\documentclass[12pt,preprint]{aastex}

\lefthead{M.~D.~Caballero-Garc\'{\i}a et~al.}
\righthead{GRO~J1655$-$40}

\begin{document}

\title{The high energy emission of GRO~J1655$-$40 as revealed with INTEGRAL spectroscopy of the 2005 outburst}

\author{M.~D.~Caballero-Garc\'{\i}a\altaffilmark{1},
        J.~M.~Miller\altaffilmark{2},
        E.~Kuulkers\altaffilmark{3},
        M.~D\'{\i}az Trigo\altaffilmark{3},
        J. Homan \altaffilmark{4},
        W.~H.~G. Lewin \altaffilmark{4},
        P.~Kretschmar\altaffilmark{3},
        A.~Domingo\altaffilmark{1},
        J.~M.~Mas-Hesse\altaffilmark{5},
        R.~Wijnands\altaffilmark{6},
        A.~C. Fabian\altaffilmark{7},
        R.~P. Fender\altaffilmark{8,6},
        M.~van der Klis\altaffilmark{6}
        }

\email{mcaballe@laeff.inta.es}

\altaffiltext{1}{LAEFF-INTA, P.O. Box 50727, 28080 Madrid, Spain, mcaballe@laeff.inta.es}
\altaffiltext{2}{Department of Astronomy, University of Michigan, 500 Church Street, Ann Arbor, USA, MI 48109, jonmm@umich.edu}
\altaffiltext{3}{ESA/ESAC, Urb. Villafranca del Castillo, PO Box 50727, 28080 Madrid, Spain}
\altaffiltext{4}{MIT Kavli Institute for Astrophysics and Space Research, 77 Massachusetts Avenue, Cambridge, MA 02139, USA}
\altaffiltext{5}{CAB (CSIC-INTA) P.O. Box 50727, 28080 Madrid, Spain}
\altaffiltext{6}{Astronomical Institute ``Anton Pannekoek'', Kruislaan 403, University of Amsterdam, Amsterdam, 1098 SJ, The Netherlands}
\altaffiltext{7}{University of Cambridge, Institute of Astronomy, Cambridge CB3 OHA, UK }
\altaffiltext{8}{School of Physics and Astronomy, University of Southampton, Highfield, Southampton, SO17 1BJ, UK}

\label{firstpage}

\begin{abstract}

We present broadband (3$-500$\,keV) {\rm INTEGRAL} X-ray spectra and
X-ray/optical light curves of the luminous black hole X-ray transient
and relativistic jet source GRO~J1655$-$40. Our analysis covers four
Target of Opportunity observations of the outburst that started in
February 2005. We find that the high energy emission of GRO~J1655$-$40
can be modelled well with an {\it unbroken} power-law (with photon
indices of ${\Gamma}=1.72{\pm}0.03,2.21{\pm}0.04$ for the first and
the second observations, respectively). These correspond to hard and
thermal dominant states, respectively. In contrast to many other
black hole spectra, high energy complexity in the form of a break or
cut-off is not required for the hard state, contrary to previous 
expectations for this state. We show for the first time that 
Comptonization by non-thermal electrons is the dominant process for 
the high energy emission in the hard state. We discuss our results in
terms of models for broad-band emission and accretion flows in
stellar-mass black holes. 

\end{abstract}


\section{Introduction} \label{introd}

GRO~J1655$-$40 is a black hole X-ray binary whose parameters are well
known.  Also called Nova Scorpii 1994, the source was discovered with
the Burst and Transient Source Experiment (BATSE) on board the Compton
Gamma-Ray Observatory (CGRO) on 1994 July 27 \citep{zhang94}. The
optical counterpart was discovered soon after by \citet{bailyn95a}
($V{\sim}14.4$ mag). Subsequent optical studies regarding the
properties of the light curve during the outburst and quiescent period
showed that the system is an LMXB composed of a blue subgiant
(spectral type F4 IV) as the secondary and a black hole (hereafter BH)
as the primary ($m_{\rm BH}=7.02{\pm}0.22$$M_{\odot}$,
\citet{orosz97a}). The system is located at a distance of 3.2 kpc as
measured by \citet{tingay95}. Although \citet{foellmi06} have recently
suggested a smaller distance, their parameters imply that the
donor star would not fill its Roche lobe.  \citet{bailyn95b}
established the orbital inclination of the system to be
${\simeq}70^{\circ}$ (see also \citet{orosz97b} and \citet{hooft98});
an independent determination made by \citet{kuulkers98} based on X-ray
flux dips constrained the inclination of the system to be
$60^{\circ}-75^{\circ}$. The inclination of the inner disk may be as
high as $85^{\circ}$, indicating a slight mis-alignment with the
binary system \citep{hjellming95}.

GRO~J1655$-$40 has displayed some of the most extreme behavior and
phenomena yet observed from any black hole X-ray transient.
\citet{strohmayer01} discovered a pair of high-frequency
quasi-periodic oscillations (QPOs) at $300$ and $450$ Hz in power
spectra from the 1996/1997 outburst of the source. If the higher
frequency is associated with the Keplerian frequency at the innermost
stable circular orbit (ISCO, \citet{shapiro83}, located at
$R_{ISCO}=6~R_{g}$ for a Schwarzschild black hole or at $R_{ISCO}=
1.25~R_{g}$ in the case of a maximal Kerr BH of $a=0.998$; where $R_{g}=GM/c^{2}$
is the gravitational radius and $a=cJ/GM^{2}$; see \citet{bardeen72}
and \citet{thorne74}), the frequency observed indicates that
GRO~J1655$-$40 harbors a spinning black hole. This suggestion is
broadly consistent with spin estimates based on fits to skewed Fe~K
emission lines ($a \geq 0.9$ as per $r \leq 1.4~R_g$,
\citet{miller05}). GRO~J1655$-$40 has also ejected extremely
relativistic radio jets \citep{hjellming95}. Finally, unbroken
power-law emission (i.e.\ without a cutoff) from GRO~J1655$-$40 has
been detected out to 800\,keV \citep{tomsick99}, offering a crucial
insight on high energy processes in black hole systems.

High energy processes and the periods of correlated behavior known as
``states'' in black hole binaries are the focus of this paper.  Three
active states are commonly recognized in the soft X-ray domain (see
\citet{remillard06}): the non-thermal dominant or hard state (formerly
called low/hard), the thermal-dominant state (formerly called
high/soft), the steep power law state (hereafter SPL). Additionally 
there are transitions between these states which
are often referred to as intermediate states. In the hard state, the
soft X-ray emission is very weak and the spectrum is dominated by some
kind of non-thermal emission that is broadly consistent with a power
law (hereafter PL) at higher energies (${\gtrsim}20$\,keV). It has
been known for a long time that black hole spectra in the hard state
are exponentially cut off at ${\sim}100$\,keV  (see \citet{grove98b}
and references therein). A radio jet is usually inferred in this state
from flat radio spectra (see \citet{fender04} for an unified model for
X-ray states and radio emission in black hole X-ray transients). In
the thermal-dominant state, the disk dominates the X-ray emission.
Although high energy emission  (${\gtrsim}20$\,keV) also is seen in
this state, it is weaker, generally steeper, and extends up to 800\,keV
without any break \citep{grove98b}. In the SPL, these emission 
components are combined -- both the disk and the non-thermal 
power-law are strong, although the disk no longer follows the 
$L{\propto}T^4$ relation that is
observed in the soft state \citep{kubota01}; \citep{saito07}. 

Several models have been proposed to explain the physical conditions
of the innermost accretion flow and the nature of hard X-ray emission
in the various states, but a clear picture has not yet been achieved. 
At low mass accretion rates hard X-ray emission may arise in an inner
region filled by a hot ($kT_{e}{\sim}100$\,keV), radiatively
inefficient, advection-dominated accretion flow (ADAF)
(\citet{narayan96}, \citet{esin01}). A recent study made by
\citet{yuan04} shows that the ADAF scenario is not able to explain the
relatively high luminosities that have been observed in the hard state
of some black-hole X-ray binaries.  Alternatively, some recent models
suggest that direct synchrotron emission, and/or synchrotron
self-Comptonization in a jet may dominate the hard X-ray emission
(\citet{markoff01}, \citet{markoff03}, \citet{markoff05}). 
Both at low and high mass accretion rates thermal Comptonization 
in a corona (fed by seed photons from the disk) may also be an 
important source of hard X-ray emission (see, e.g., \citet{frontera03}). 
An alternative source of Comptonization which is less reliant on 
the disk is  bulk motion Comptonizing (BMC, \citet{ebisawa96}; 
see also \citet{titarchuk02} for a co-moving Comptonizing medium).  
In this case, the Comptonization is due to bulk motion of an 
almost free falling (convergent accretion) flow close to the black hole.

A thermal distribution of the Comptonizing particles (electrons) 
necessarily leads to a turnover in the emitted spectrum around $kT$; 
therefore, thermal Comptonization should lead to a turnover near to the electron
temperature of the corona, $kT_{e}$. Synchrotron emission and
non-thermal electron distribution do not necessarily predict
such a turn-over, however, and this difference provides an
observational tool to distinguish which processes dominate the hard
X-ray emission in different black hole states. The high energy
sensitivity of INTEGRAL is especially well-suited to this purpose.

In this paper, we report on observations of GRO~J1655$-$40 made with
INTEGRAL during the outburst which began in February of 2005 
\citep{markwardt05}. In Section \ref{observ} we describe our
observations and in Section \ref{curves} we show the light curves
obtained with JEM-X, ISGRI, SPI and OMC (instruments on-board
INTEGRAL) and RXTE and discuss the possible origin of their evolution.
In Section \ref{spectext} we present our spectral analysis made with
JEM-X, ISGRI and SPI. Finally, in Section \ref{discuss} we discuss our
results in the context of the different models and theories present in
the literature.

\section{Observations} \label{observ}

\subsection{INTEGRAL observations}

The data were obtained with INTEGRAL and cover the first part of the
2005 outburst using the following instruments: the SPectrometer on
INTEGRAL (SPI; \citet{vedrenne03}), the INTEGRAL Soft Gamma-Ray Imager
(ISGRI; \citet{lebrun03}), the Joint European X-ray Monitor (JEM-X;
\citet{lund03} and the Optical Monitoring Camera (OMC;
\citet{mas-hesse03}). ISGRI is optimized for 15\,keV to 10 MeV imaging
and SPI is optimized for high-resolution spectroscopy in the 18\,keV
to 8 MeV band. The former provides an angular resolution of $12'$
full-width half maximum (FWHM) and an energy resolution, $E/{\Delta}E$
of ${\approx}12$ (FWHM) at 100\,keV.  SPI provides an angular
resolution of $2.8^{\circ}$ (FWHM) and an $E/{\Delta}E$ of $430$ FWHM
at 1.3 MeV. JEM-X has a fully coded Field of View (FOV) of
$4.8^{\circ}$ diameter and an angular resolution of $3'$ FWHM. JEM-X
has medium resolution spectral capabilities in the energy range of
3-35\,keV.  The OMC is an optical monitor, with a FOV of
$5^{\circ}{\times}5^{\circ}$ and an astrometric resolution of $<1$
arcsec, and performs optical photometry in the V-band down to $18^{th}$
magnitude.

Our program consisted of 4 ToO observations of 100 ks each, spread
from 27 February to 11 April of 2005 (we will refer to these as epochs
1--4 below; see Table \ref{tobserv} for more details). The difference
in the exposure times between the INTEGRAL instruments given in Table
\ref{tobserv} (JEM-X, ISGRI and SPI) are due to the difference in the
dead times and variation in the efficiency along the fields of view.
The dithering pattern used during the observations was $5{\times}5$
(square of 25 pointings separated by 2.17 degrees centered on the main
target of the observation); this is the best pattern in order to
minimize background effects for the SPI and ISGRI instruments in
crowded fields. Data reduction (in the case of JEM-X, ISGRI and SPI)
was performed using the standard Off-line Science Analysis (OSA) 5.1
software package available from the INTEGRAL Science Data Centre
(ISDC; \citet{courvoisier03}). In the case of SPI, because of the
lower angular resolution and crowded field of view in ${\gamma}$-rays
at this position of the sky
($l$,$b$)=($344.98^{\circ}$,$+2.46^{\circ}$), (see Figure
\ref{mosaic}), we used a non-standard procedure in the analysis of the
data, described in \citet{deluit05} and \citet{roques05}. For the same
reason, in the case of OMC, we used a standard pipeline available in
OSA 6.0 (recently delivered) for extraction of fluxes. For  OMC all
the public data available from ISDC were downloaded  (this resulted
only in a slight increase in the ammount of data).  Because of the
steep fall in response in the case of JEM-X,  and because of its
reduced FOV (${\approx}5^{\circ}$ of diameter), we limited the
pointing radius with respect to the GRO~J1655$-$40 position to be
within $4^{\circ}$. In the case of SPI and ISGRI, with large Fully
Coded Fields of Views (FCFOV) ($16^{\circ}{\times}16^{\circ}$ for SPI
and $8.3^{\circ}{\times}8^{\circ}$ for ISGRI), pointing selections
were not necessary. In total, 199 individual pointing (or Science
Window -- each having exposure times lasting from 1800 to 3600 s and
following a $5{\times}5$ dithering pattern on the plane of the sky
\citet{courvoisier03}) data were used for both SPI and ISGRI, 96
pointings for JEM-X and 66 pointings for OMC.

\subsubsection{Extraction of light curves}

GRO~J1655$-$40 was covered with INTEGRAL as part of the Galactic
Bulge Monitoring Program \citep{kuulkers07}. Precisely at the start of
this program  GRO~J1655-40 was reported to become active
\citep{markwardt05}. The subsequent outburst of GRO~J1655-40 was also
followed with the RXTE/ASM and with a dense program of pointed
RXTE/PCA observations (Homan 2005). In Figure \ref{lcurves} we show
the ISGRI (20-60 and 60-150\,keV), RXTE/ASM (2-12\,keV)  and OMC
(optical) light curves. The light curve derived from the IBIS/ISGRI in
the 20-60 and 60-150 keV  energy bands had 1\,800 sec exposures (about
150 counts/s correspond to 1 Crab on-axis in the 20-60\,keV energy 
band, see Appedix A in \citet{kuulkers07}). GRO J1655-40 was observed
at a large off-axis angle (${\approx}15^{\circ}$  from the center of
the field of view; so is in the partially coded field of  view of
IBIS/ISGRI and not visible with X-ray monitor JEM-X).

Since GRO~J1655$-$40 is located in a very crowded FOV for the OMC, in
the flux extraction process we force the photometric aperture to be
centered at the source coordinates, which are taken from the OMC Input
Catalogue \citep{domingo03}.

The typical limiting magnitude of the OMC in the Galactic bulge is
between $V=15$ and $V=16$ ($3{\sigma}$). This value depends strongly
on sky background and source contamination. In our case
($l$,$b$)=($344.98^{\circ}$,$+2.46^{\circ}$) we can confidently get
limiting magnitudes down to $V=16$ ($3{\sigma}$). For each INTEGRAL
pointing, the OMC monitors the sources in its FOV by means of shots of
variable integration time. Typical values in the range 10 to 200\,s
(currently 10, 50 200\,s) are used to optimize sensitivity and to
minimize read-out noise and cosmic-ray effects. For the faintest
objects, several 200\,s exposures in the same pointing can be combined
during data analysis on the ground. We obtained one photometric point
by combining several 200\,s OMC shots.  To increase the signal to
noise ratio, we combined the individual photometric points of every
pointing and calculated the final photometric points by making running
averages in order to minimize the dispersion.

\subsubsection{Extraction of spectra}

For JEM-X and ISGRI, individual spectra were obtained for each
pointing. The spectra were then combined to obtain an averaged
spectrum per epoch using the $spe\_ pick$ OSA tool and standard
procedures (see \citet{jemx05} and \citet{ibis05}) to rebin the
response matrices.  In the case of SPI, we derived directly one
spectrum per revolution. The SPI spectra were extracted over an energy
range of (23-800)\,keV (with 26 logarithmic bins) using the SPIROS
package within OSA, applying maximum likelihood optimization
statistics \citep{skinner03}. Due to the crowdedness of the field and
the low spatial resolution of the instrument (i.e. sources separated
less than $2.8^{\circ}$ can not be resolved) it was necessary to apply
special techniques to extract spectra (see \citet{roques05} and
\citet{deluit05}). This included accounting for variability of the
sources present in the FOV.  The background \footnote{Due to the fact
that JEM-X, ISGRI and SPI are detectors based in coded mask optics,
the detection of sources is made in basis of a deconvolution process,
taking also into account background, in an iterative process called
IROS (Iterative Removal Of Sources). In OSA 5.1 the spectra obtained
are always background subtracted and it is not necessary to apply in
XSPEC any background correction to the obtained spectra. The light
curves obtained are also background subtracted. However, in JEM-X
analysis it is possible to skip the level of background
subtraction. We refer to the \citet{osaspi}, \citet{goldwurm03},
\citet{westergaard03} and \citet{dubath05} for more details.}  was
determined by using flat fields from particular INTEGRAL revolutions
(revolution 220 in our case -- i.e. the time closest publicly available
flat field to our period of observations), and the use of background method $3$
(determination of the background based on some specific flat field
INTEGRAL observations). Due to the fact that SPI is a high resolution
spectrograph in its energy range (20-8\,000\,keV) with a reduced
number of detectors (namely, 19) it is not optimized in the detection
of sources without taking into account previous information about 
the spatial distribution of them. Thus, we used as an input catalog 
of sources those obtained from the ISGRI in the (20-40)\,keV mosaic images 
(see Figure \ref{mosaic}).

The signal from GRO~J1655$-$40 was too soft to detect any emission in
the last two epochs with SPI. We combined single revolution SPI
spectra from each epoch, since there was no significant evolution and
in order to increase the signal to noise ratio. This resulted in one
SPI spectrum in revolution 290 and another one combining the
revolutions 295 and 296. The same was done for the low energy
instruments (JEM-X and ISGRI), thus obtaining four spectra, namely one
for each epoch, as can be seen in Table \ref{tobserv}. We applied
2$\%$ systematic errors to the JEM-X, ISGRI and SPI spectra. We
restricted our analysis in the energy ranges of $5-30$, $23-600$ and
$23-800$\,keV for the JEM-X, ISGRI and SPI spectral analysis, as
recommended in \citet{lubinski05}. The SPI and ISGRI spectra were 
rebinned at high energies (${\gtrsim}200$\,keV) with the FTOOL 
$grppha$ procedure to reach the detection level of $3{\sigma}$.

\section{Analysis of light curves} \label{curves}

Figure \ref{lcurves} shows the GRO~J1655$-$40 light curves obtained
with ISGRI (from the INTEGRAL Galactic Bulge Monitoring Program;
\citet{kuulkers07}) in two energy bands (60-150\,keV and 20-60\,keV)
together with the OMC (optical).  The public RXTE/ASM (2-12\,keV)
light curve in the same period of time is shown in the same Figure for
comparison.

The light curve in Figure \ref{lcurves} does only show the first month
of outburst of GRO~J1655$-$40.  For a full outburst light curve, refer
to  \citet{brocksopp06}. \citet{brocksopp06} show the X-ray light
curve  obtained by SWIFT, jointly with that obtained in the optical
and ultraviolet band using the Ultraviolet/Optical telescope on-board
SWIFT (UVOT). In Figure \ref{lcurves2} we show the RXTE/PCA light
curve in the range (2-60)\,keV plus a hardness ratio  (calculated as
the ratio of the 9.4-18.2\,keV and 2.8-5.7\,keV count rates) evolution
of the source during the entire outburst, taken from \citet{homan07}.
The light curves were made from 520 RXTE observations, with one
(averaged) data point for each observation. The horizontal lines shown
in Figures \ref{lcurves} and \ref{lcurves2} indicate the time
intervals (100 ks each; in 4 observations) over which our average
spectra were obtained. Note that black hole  transients usually begin
and end their outbursts in the hard state (see \citet{nowak95},
\citet{fender04}, \citet{homan05}) and the 2005 outburst of
GRO~J1655--40 is no exception. As can be seen in Figure 
\ref{lcurves}, the beginning of the outburst first started at high
energies ($20-150$\,keV) rather than at softer X-rays ($2-10$\,keV).
Moreover, as can also be seen from the hardness ratios in Figure
\ref{lcurves2}, the observations of epoch 1 correspond to the hard
state of black holes (see Section \ref{discuss}). The other three
epochs were done during a period when the source spectrum was much
softer. However, it should be noted that it was only later in the
outburst that the softest spectra (corresponding to the
thermal-dominant state) were observed (see Figure \ref{lcurves2}).

The first indications of an impending outburst of GRO~J1655-40 came
from RXTE/PCA bulge-scan observations on 2005 February 19
\citep{markwardt05}. On February 20, observations made with  the PANIC
camera on the 6.5\,m Magellan-Baade telescope at Las Campanas 
Observatory revealed a J-band (near-infrared) magnitude of
J=13.2${\pm}$0.1 \citep{torres05}, indicating that GRO~J1655-40 was
brighter by ${\approx}$0.5 mag in J relative to its magnitude
in quiescence  (J=13.7-14) \citep{greene01}. On the same date a radio
detection of the source was reported \citep{rupen05}.

As can be seen from \citet{buxton05} and \citet{brocksopp06} the
optical light  curve behaves differently from the X-rays. The optical
behaviour  consists of an increase in the flux until a constant
level at MJD=53\,455. This behaviour can also be seen with the
OMC, because the flux in the optical light curve was increasing,
reaching a constant and detectable value ($V{\sim}15$ mag) in
MJD=53\,455. Then the optical flux became constant. It is also
interesting to  note the rapid increase of the radio emission
at MJD${\approx}$53\,450, coinciding with our epoch 3 of
observations and also with the beginning of the plateau in both the
optical and infrared light curve.

\section{Spectral analysis of INTEGRAL data} \label{spectext}

We performed fits to the combined JEM-X, ISGRI and SPI spectra, for
each of the four epochs (see Table \ref{tobserv}) using XSPEC
\citep{arnaud96} v.11.3. All errors quoted in this work are 90\%
confidence errors, obtained by allowing all variable parameters to
float during the error scan.  In all fits, we fixed the value of the
column density to $N_{H}=8.0{\times}10^{21}$ atoms$/cm^{2}$ as
obtained by \citet{diaz06} using XMM-Newton data of GRO~J1655$-$40
during the same outburst.

Our main aim in these fits is to characterize the broad continuum as
seen with INTEGRAL and its broad energy coverage. To account for 
uncertainties in relative instrument calibrations, we fixed JEM-X
multiplicative calibration constant to 1 and let that of ISGRI and SPI
free to vary in the fit for the different data sets as shown in 
Table \ref{param_spec}.

In Figures \ref{fspec1}, \ref{fspec2}, \ref{fspec3}, \ref{fspec4} and
\ref{fspec5}
we show spectra from each epoch because these provide a 
convenient way to see the evolution of
the source during our observations.  We fitted the spectra with
several models and the derived parameters are presented in Table
\ref{param_spec}.  We modelled the spectra with simple and
phenomenological disk plus power-law models and common Comptonization
models described in Section \ref{introd}.  We find that the former
are very successful (see Section \ref{powerlaw}), and there is no evidence for
any spectral break in the data up to ${\approx}600$\,keV
(see Section \ref{compttmod}).  Although a simple power-law provided
marginally acceptable fits in the first two epochs, it was necessary to include
an iron line and edges components in the fits of the last two epochs
(see Section \ref{edges}).

\subsection{Fits of the former two epochs}

\subsubsection{Fits with a pure power-law} \label{powerlaw}

We initially performed fits with a phenomenological power-law model
($powerlaw$ model in XSPEC) in the spectra that extend up to
$600$\,keV.

In the first epoch, the source spectra extend in the energy range of
$5-20$\,keV for JEM-X and $23-600$\,keV for both ISGRI and SPI,
respectively. In fitting with the $power-law$ model, we obtained a
reduced chi-square of ${\chi}^{2}_{\nu}=1.26$ with ${\nu}=63$ (${\nu}$
meaning the number of the degrees of freedom of our fit). The presence
of a multicolor disk component was not significantly required in our
fit. The value obtained for the photon index was 
${\Gamma}=1.72{\pm}0.03$, common for black holes in a hard spectral
state.

For the second epoch of our observations, the source spectra extend in
the energy range of $5-30$\,keV, $23-500$\,keV and $23-400$\,keV for
JEM-X, ISGRI and SPI, respectively. This time the presence of emission
coming from a disk was significant, and we added an absorbed multicolor
disk component \citep{mitsuda84} to the power-law ($phabs(diskbb + powerlaw)$ 
model in XSPEC). We obtained ${\chi}^{2}_{\nu}=1.38$ with ${\nu}=88$. 
The value obtained for the inner disk temperature was
$kT_{in}=1.25{\pm}0.01$\,keV. The spectrum is softer,
with a value for the photon index of ${\Gamma}=2.21{\pm}0.04$.

For the last two epochs, in fitting with a phenomenological power-law
(plus a multicolor disk black body component), the spectra showed
large negative residuals at ${\approx}10$\,keV, compatible with
the presence of $Fe$ edges.  The presence of
positive residuals in the range $6-8$\,keV is consistent with the
presence of a broad $Fe$ emission line. In Section \ref{edges} we go
into more detail on the fits of both these epochs.

\subsubsection{Fits with thermal Comptonization models} \label{compttmod}

As explained in Section \ref{introd}, all models involving thermal
Comptonization processes as the origin of the high energy emission
seen in black hole transient systems predict an energy turn-over at
the electron temperature of the corona.  So, in order to assess the
role of Comptonization during the time of our observations, we fitted
with the thermal Comptonizing model of \citet{titarchuk94} (called
$CompTT$ in XSPEC), which deals with the special case of high
temperatures and/or small opacities (so the relativistic effects are
taken into account).  We also made fits using the BMC model of
\citet{ebisawa96} (see also also \citet{titarchuk02} for a co-moving
Comptonizing medium), which deals with the Comptonization due to bulk
motion of the almost free falling (convergent accretion) flow close to
the black hole.  Moreover, we fitted with a phenomenological 
and multiplicative cutoff (called $highecut$ in XSPEC) model in 
order to find any turn-over energy in our data which could be in 
agreement with Comptonization models playing the major role in the 
high energy emission of GRO~J1655$-$40.

We thus fitted the spectra of the first two epochs \footnote{As we
explained in Section \ref{powerlaw} the presence of significant
residuals compatible with broad $Fe$ emission line, edges and likely
reflection could distort our undertanding of the continuum. Thus, we
do not use last two epochs in our study of the continuum
models. Moreover, the spectra in these epochs do not show significant
emission at high (${\gtrsim}150$\,keV) energies.}  with following
models: $phabs(diskbb+powerlaw)highecut$ (hereafter called model 1),
and $phabs(diskbb+compTT)$ (hereafter called model 2) and $phabs(bmc)highecut$ 
(hereafter called model 3). In the case of the first epoch, the 
presence of a multicolor disk component was not significantly required 
in our fit. We assumed a spherical geometry for the Comptonizing medium 
for the model 2 in all our fits. 

For epoch 1 of our observations, we obtained the following statistics for  all
three models, i.e. ${\chi}^{2}_{\nu}=1.13,1.17,1.13$
(${\nu}=61,60,59$), respectively. We find that when fitting the data
with a cut-off power-law (model 1), while showing slightly better
statistics,  the high energy cut-off can not be well constrained
(values between 5 and 54 keV are compatible with the data). The
obtained folding energy is $>253$\,keV. We made fits to separate {\rm
JEM-X+ISGRI} and {\rm JEM-X+SPI} to see if  in one of the two data
sets a break is possible. In the former, we obtained  uncostrained
values for the cut-off and the folding energies, being in the ranges
of 5-38\,keV and $>238$\,keV, respectively. In the second data set
there was  not any break in the data as well. We conclude that the
cut-off features  found are of instrumental origin and are not
physically meaningful. Thus, there is not any real cut-off in our
data. This issue is also supported by the  fact that the parameters
optical depth and temperature of the electrons (${\tau}$ and $kT_{e}$)
when fitting with model 2 are unconstrained as well. This model
clearly is not a good description of the high energy spectrum
(${\gtrsim}20$\,keV). The model 3, while giving the lowest residuals, 
does not allow to constrain the values for the parameters (neither of 
$bmc$ and $highecut$ components). Thus, this model does not represent
a  physical description of the data. The fact that model 3 provides
slightly better  statistics than the power-law model described in
Section \ref{powerlaw}  would be due to the fact that this model is a
convolution of both soft and hard emission components taking into
account the physical condition in the inner region of the accretion
disk. We derive from fitting with all these models that high energies
can not be reproduced by thermally  upscattering photons by a
Comptonizing corona alone. In order to take this issue into account,
we fitted the spectrum of this epoch with the hybrid Comptonization
model EQPAIR \citep{coppi99} proper for very hot
plasmas which describes the physics and emission properties of hybrid
plasmas, where the particle distribution energy is approximately a 
Maxwellian plus a power-law. This model deals a hot plasma cloud, mainly
modeled as a spherical corona around the compact object, illuminated 
by soft thermal (Maxwellian) and
non-thermal (either power-law or monoenergetic distributed) electrons 
that lose energy by Compton, Coulomb and bremsstrahlung interactions. 
This model was shown to be successful in accounting for the high energy
spectra of Cygnus X-1 and other black hole candidates in different 
spectral states and over a broad energy band ranging from soft X-rays 
to gamma-rays (see e.g. \citet{mcconnell00}, \citet{mcconnell02},
\citet{cadolle06}, \citet{malzac06}). 

For epoch 2 of our observations, we obtained the following statistics for all
three models, i.e. ${\chi}^{2}_{\nu}=1.47,1.45,2.07$
(${\nu}=86,87,86$), respectively.  Again, model 1 does not show a
break or cut-off below $500$\,keV (i.e. values of both cut-off and
folding energies can not be  costrained) and model 3 is not a proper
fit to the data due to the bad  statistics. In this epoch, the values
obtained for the optical depth of  the Comptonizing medium and the
temperature of the electrons (${\tau}$ and $kT_{e}$) can not be
constrained. Thus, the data show that thermal  Comptonization is not
the main process that generates the emission at high energies
(${\gtrsim}20$\,keV). As in epoch 1, we tested the model of
\citet{coppi99}, optimized for very hot coronae (see description in
paragraph above), this time coupled with $diskbb$ component to
describe the soft emission from the accretion disk.

\subsubsection{Fits with the {\rm EQPAIR} non-thermal Comptonization model} \label{comptteqpair}

This model (EQPAIR) takes into account angle dependence, Compton 
scattering (up to multiple orders), photon pair production, pair 
anihilation, bremsstrahlung  as well as reflection from a cold disk. 
As noticed by \citet{coppi99}, if a spectrum extends up
to $500$\,keV high energy emission coming from a non-thermal
population  of electrons is clearly present. 

As can be seen in Figure \ref{fspecpo1} and \ref{fspecpo2}, in which we
fitted both 1 and 2 epochs with a single power-law (see description
in Section \ref{powerlaw}), an small deficit of counts of the ISGRI spectra
above 200\,keV with respect to SPI is clearly present. Moreover, the 
large bins of ISGRI spectra in this range of the spectra, would indicate
that the source is not detected above 200\,keV. This is a comprehensive
issue since SPI instrument is optimized for doing spectroscopy 
up to 1\,MeV while ISGRI has poorer sensitivity above 200\,keV.
Taking these considerations into account, we limited our analysis of
ISGRI data in the energy range of $23-200$\,keV.

The $eqpair$ model allows to inject a non-thermal electron distribution
with Lorentz factors between ${\gamma}_{min}$ and ${\gamma}_{max}$ and
a power-law spectral index ${\Gamma}_{p}$. The cloud is illuminated
by soft thermal photons emitted by an accretion disk. These photons
serve as seed for Compton scattering by both thermal and non-thermal
electrons. The systems is characterized by the power (i.e. luminosity)
$L_{i}$ supplied by its different components. We express each of them
dimensionlessly as a compactness parameter, 
${\rm l}_{i}={\rm L}_{i}{\sigma}_{\rm T}/({\rm R}{\rm m}_{e}{\rm c}^{3})$,
where ${\rm R}$ is the characteristic dimension and ${\sigma}_{\rm T}$
the Thompson cross-section of the plasma. Thus, ${\rm l}_{s}$, ${\rm l}_{th}$,
${\rm l}_{nth}$ and ${\rm l}_{h}={\rm l}_{th}+{\rm l}_{nth}$ correspond to
the power in soft disk entering the plasma, thermal electron heating,
electron acceleration and the total power supplied to the plasma.
The total number of electrons (not including ${\rm e}^{+}$
and ${\rm e}_{-}$ pairs) is determined by ${\tau}$, 
the corresponding Thompson optical depth, measured from the center 
to the surface of the scattering region. We considered the source to 
be moderately compact and fixed ${\rm l}_{s}=10$, as broadly reported
for other sources with similar characteristics. 

The disk spectrum incident on the plasma is modelled with a multicolor
disk blackbody as given by the $diskbb$ model in XSPEC of \citet{mitsuda84}.
The temperature of the inner edge of the accretion disk was fixed to
${\rm kT}_{e}=0.5$\,keV. The limits of the accretion disk were fixed
to ${\rm R}_{max}=100{\rm R}_{g}$ and ${\rm R}_{min}=6{\rm R}_{g}$. We 
attempted to fit both spectra of epochs 1 and 2 fixing the reflection 
covering factor to zero.

For epoch 1 of our observations we first performed a fit with non-thermal 
electrons injected with a power-law distribution of Lorentz factors
from ${\gamma}_{min}=1.3$ to ${\gamma}_{max}=1000$. The upper and the 
lower limits ${\gamma}_{min}$ and ${\gamma}_{max}$ were kept fixed 
while fitting the power-law index ${\Gamma}_{p}$. This resulted in
an acceptable fit with a reduced chi-square of ${\chi}_{\nu}=1.21$
(${\nu}=57$). The unfolded broadband spectrum and residuals are shown
in Figure \ref{fspec1}. The best fit paramaters are presented in Table
\ref{param_spec}. ${\rm l}_{h}/{\rm l}_{s}$ is about unity, i.e. 
intermediate between what is found in the hard state (4-10)
and the thermal dominant state (${\leq}0.4$) \citep{ibragimov05}. The 
heating of the plasma is dominated by the non-thermal acceleration
(${\rm l}_{nth}/{\rm l}_{h}{\approx}1$). We also fitted the spectra
of this epoch considering a mono-energetically distributed population
of non-thermal injected electrons instead. The obtained reduced chi-square 
is slightly better (${\chi}_{\nu}=1.19$, ${\nu}=57$). The values obtained
for ${\rm l}_{h}/{\rm l}_{s}$ and ${\rm l}_{nth}/{\rm l}_{h}$ remained 
unchanged, except for an increasing of the Thompson scattering depth
(from $1.0{\pm}0.6$ to $2.7{\pm}0.4$, for the former and the second fit,
respectively). The increasing of the optical depth gives support to
the scenario of Comptonization through injected non-thermal electrons
as being the dominant mechansim in the sense that if this is done by 
electrons with a mono-energetic distribution then a denser cloud would 
produce the same effect as taking into account a broader distribution 
(both in energy and spatially). Thus, we conclude that in order to
reproduce the spectrum of first epoch, we have to consider an almost 
purely distribution of non-thermal accelerating particles.

As in epoch 1, we tested the model of \citet{coppi99}, 
optimized for very hot coronae in observations of epoch 2, this time coupled 
with $diskbb$ component to describe the soft emission from 
the accretion disk. We performed a fit with non-thermal
electrons injected with a power-law distribution of Lorentz factors
from ${\gamma}_{min}=1.3$ to ${\gamma}_{max}=1000$. The upper and the
lower limits ${\gamma}_{min}$ and ${\gamma}_{max}$ were kept fixed
while fitting the power-law index ${\Gamma}_{p}$. This resulted in
an acceptable fit with a reduced chi-square of ${\chi}_{\nu}=1.44$
(${\nu}=83$). The unfolded broadband spectrum and residuals are shown
in Figure \ref{fspec2}. The best fit paramaters are presented in Table
\ref{param_spec}. ${\rm l}_{h}/{\rm l}_{s}$ again is about unity.
However, the heating of the plasma by non-thermal particles is practically zero 
(${\rm l}_{nth}/{\rm l}_{h}{\approx}0$). The situation does not
improve by considering a mono-energetic distribution of the non-thermal
accelerated electrons. This issue is not surprising, since, as claimed
by \citet{coppi99}, a very good spectrum extending above 200\,keV is mandatory
in order to disentangle a likely population of non-thermal particles
in the source. In the case of our observations, the large bin at
${\approx}200$\,keV shows that the source was not detected above
this energies. We conclude that while in epoch 2 of our observations
a break is not observable in the high energy data ${\gtrsim}20$ the
energy coverage is not great enough in order to test the presence 
of non-thermal processes. This is due to the fact that in this period
the spectrum became to be very soft, compared to epoch 1, with
radiation detected to ${\approx}200$\,keV as maximum.

\subsection{Fits to epochs 3 and 4} \label{edges}

\subsection{Fits with $Fe$ emission line and absorption edges}

As shown in the shape of the residuals (see
Figure \ref{fspec3}), fits with simple multicolor disk model of
\citet{mitsuda84} and power-law did not give formally acceptable fits
to the last two epochs.  The presence of large residuals at $6-8$\,keV
and around 10\,keV require us to take into account iron emission line,
iron absorption edges and likely disk reflection components. These
features are theoretically required in very ionized mediums as the
close vicinity of the black holes (see \citet{ueda98} for a former
detection of $Fe$ absorption lines and edges in GRO~J1655$-$40 in the
context of observations of the 1996 outburst and \citet{george91} and
\citet{laor91} for a description of the $Fe$ line profile produced in
accretion disks around black holes).  The reflection component takes
into account the physical condition of the Compton-reflected continuum
of a source of hard X-rays by an accretion disk
\citep{gilfanov00}.  

We fitted our spectra with the following model: $phabs(diskbb +
powerlaw + gaussian)edge*edge$ and the results of the fits are shown
in Table \ref{param_spec}. Regarding the Gaussian $Fe$ emission line
component, we constrained the line center in the range of
$6.4-6.97$\,keV, which is the allowed range due to the different
ionization states of $Fe$.  Also, we constrained the width
(${\sigma}$) of the iron emission line to be 1\,keV as a fiducial
maximum value in order to get convergence of our fits. Regarding the
two iron absorption edges, one was fixed at 9.278~keV, which corresponds
to Fe XXVI and is expected to appear given the likely range of
temperatures and ionization.

Fitting this model to the spectra for epochs 3 and 4, we obtained
better results than fitting with a power-law model (plus an absorbed
multicolor disk), i.e. ${\chi}^{2}_{\nu}=9.99$, with ${\nu}=21$, and
${\chi}^{2}_{\nu}=2.73$, with ${\nu}=22$, for the third and fourth
epoch, respectively. Although the fit was unacceptable for the third
epoch, because of the large residuals at around ${\approx}10-20$\,keV,
which could be due to reflection \footnote{It is important to notice
that \citet{diaz06} did not find reflection signatures in the analysis
of joint XMM and INTEGRAL data corresponding to our fourth epoch. The
disagreement could be due to the $20-30$\,keV removed bin in their
study, made in order to improve the statistics for the joint spectrum
obtained in this period.}, the fit was reasonable for fourth
epoch. The multicolor disk component gave an inner disk temperatures
of $kT_{in}=1.27{\pm}0.17$\,keV for the fourth epoch, in which we also
obtained a very soft photon index (${\Gamma}= 4.7{\pm}0.6$).

We substituted the $power-law$ component by the $pexriv$ reflection
model in the third epoch. This component is a power-law with an
exponential cut-off in order to take into account reflection effects
in the spectrum. In fitting with the pexriv model, we imposed an
overabundance of $Fe$ of 2.8 with respect to the solar value and
$cos(i)=0.45$, the last implying an inclination of $63^{\circ}$ for
the inclination of the reflection medium, namely the disk (as found by
\citet{diaz06} based on XMM and INTEGRAL data for the same period of
observations). The temperature of the disk that gave the smallest
residuals was $T_{disk}=1.2{\times}10^{+07}$\,K so we fixed this value
in our fits. This value is consistent with having very ionized 
material in the accretion disk. Also, in order to properly fit the high energy part of
the spectra (${\gtrsim}20$\,keV) it was necessary to fix the $pexriv$
e-folding energy to a very high value, namely $1000$\,keV, implying a
non-detection of any cut-off up to ${\approx}200$\,keV. 

As a result of this fit for third epoch, we obtained a reflection
covering factor of $R{\leq}0.3$. Actually, this value ($R$)
approximately measures the solid angle subtended by the reflecting
medium as seen from the source of the primary radiation,
$R{\approx}{\Omega}_{refl}/2{\pi}$, so that $R=1$ for an isotropic
point source above an infinite optically thick slab. We fixed the
ionization parameter to a very high value, consistent with a very high
ionization medium, $x_{i}=4{\pi}F_{ion}/n$=$5000$\,erg${\times}$cm$/$s
(where $F_{ion}$ is the 5-20\,keV irradiating flux and $n$ is the
density of the reflector; see \citet{done01}).

In order to estimate the source luminosity in each of our
observations, we used the power-law plus an absorbed multicolor disk
model for all the epochs. The contribution of the iron emission and
absorption effects together contribute only by $<10\%$ of our data in
the last two epochs. The un-absorbed flux in the range of $5-100$\,keV
energy range is $F(5-100)_{X}=1.4{\times}10^{-9}, 1.2{\times}10^{-8},
1.0{\times}10^{-8}, 1.0{\times}10^{-8}$ erg\,cm$^{-2}$ s$^{-1}$ for
the first to fourth epoch, respectively.  For a distance of 3.2 kpc
\citep{tingay95}, these fluxes give luminosities in the $(5-100)$\,keV
energy range of $L(5-100)_{X}=1.7{\times}10^{36}, 1.5{\times}10^{37},
1.2{\times}10^{37}, 1.2{\times}10^{37}$ erg$/$s. The un-absorbed
fluxes in the 5-600 \,keV energy range are
$F(5-500)_{X}=3.7{\times}10^{-9}, 1.2{\times}10^{-8}$,
$1.0{\times}10^{-8},1.0{\times}10^{-8}$ erg\,s$^{-1}$ cm$^{-2}$,
giving luminosities of $L(5-600)_{X}=4.6{\times}10^{36},
1.5{\times}10^{37}$, $1.2{\times}10^{37}, 1.2{\times}10^{37}$ erg$/$s
for the first to fourth epochs, respectively. Even if we take the
broader energy range as more indicative of the true source luminosity,
these values represent only 0.5\%, 1.5\%, 1.3\%, and 1.3\% of the
Eddington limit ($L_{\rm EDD}=9.4{\times}10^{38}$ erg$/$s for a black
hole of 7 $M_{\odot}$), respectively.

\section{Discussion and Conclusions} \label{discuss}

We have made fits to four epochs of {\rm INTEGRAL} broad-band spectra
of the stellar-mass black hole GRO J1655$-$40 during its 2005
outburst.  We find that GRO~J1655$-$40 was in the hard state in the
first epoch, based on the low photon index (i.e. $1.72{\pm}0.03$) and
the absence of a strong thermal disk component. The source evolved to
a state that resembles a thermal dominant state in the classification
scheme of \citet{remillard06}. These statements are in agreement with 
the previous study of this outburst by \citet{brocksopp06}.  However,
\citet{saito07}, on the basis of their analysis of RXTE/PCA data of
this source, showed that the luminosity of the accretion disk deviates
from the the $L{\propto}T^{4}$ relation typical of thermal-dominant
states during our other 3 observations (epochs 2, 3 and 4).  So, the
state observed may not be a true thermal-dominant state, and/or
non-thermal effects may need to be modeled to describe the accretion
disk emission \citep{kubota01}.

For the two latest epochs, we found that our data is best fitted by
adding an iron emission line and edges in the model consisting of an
absorbed multicolor disk \citep{mitsuda84} plus a power-law component.
Although the obtained fits were not formally acceptable
(${\chi}^{2}_{red}=9.99, 2.73$ for the third to fourth epochs, 
respectively), the fit in the fourth epoch was improved by 
${\Delta} {\chi}^{2}_{red}=4.8$. Also, the shape of the residuals 
using the model consisting of a
multicolor disk black body plus a power-law showed a clear excess in
the $6-8$\,keV range and a drop at around 10\,keV, features that can
only be explained by the presence of iron emission line and the
presence of $Fe$ edges.  The spectrum in the third epoch showed an
excess of absorption in the $10-20$\,keV energy range with
respect of that expected taking into account the edges obtained in
\citet{diaz06} based on XMM and INTEGRAL spectra (see Section \ref{edges}).
This excess could be explained by the
presence of reflection, but the situation is still unclear. We
attempted to deal with this feature in third epoch by fitting with a
broken power-law model ($pexriv$ in XSPEC) but there is no clear
evidence for a break in this spectrum. We conclude that for
GRO~J1655$-$40, it is difficult to study reflection features. This may
be due to the high inclination of the source, which could alter the
shape of the reflection features through scattering.  The presence of
these features in the spectra of epochs 3 and 4 may reveal differences
in the disk outflow properties with respect to any outflow in epochs 1
and 2. In fact, our spectrum obtained in epoch 3 precedes a {\it
Chandra} observation revealing a line-rich spectrum\citep{miller06}~
by only 3 days.

Assuming a value of $63^{\circ}$ \citep{diaz06} for the inclination and 
a distance of 3.2\,kpc \citep{tingay95}, the disk
normalization factor (in fits made only with a power-law model plus a
multicolor absorbed disk) gives inner disk radii \footnote{This radii
are measured from infinity, using the formula: \\
$R_{in}=D{\times}(diskbb_{norm}/cos(i))^{0.5}$\\ Where $R_{in}$ is in
km and D is in units of 10\,kpc.  This formula is inferred without
taking into account gravitational effects from the General Relativity
Theory. If gravitational corrections would be taken into account, then
smaller co-moving radius would be obtained.} of $16.5{\pm}0.2$\,km for
the second epoch (the first epoch was in a hard state and had no
significant contribution from the disk, as shown in Section
\ref{powerlaw}). Since $R_{g}=10.4$\,km is the value of the
gravitational radius for a $7M_{\odot}$ black hole, we find that the
matter arrives at inner radii of ${\approx}1.6R_{g}$ in the second
epoch.  This value is consistent with the value predicted for a
maximally rotating black hole as explained in Section \ref{introd}
($R_{ISCO}=1.25~R_{g}$) and is similar to that obtained by
\citet{tomsick99}.  Analyzing {\it RXTE} data, \citet{tomsick99} found
inner radii values of
$R_{in}=10.9^{+2.6}_{-2.6}-21.9^{+5.2}_{-5.2}$\,km, depending on the
value adopted for the disk inclination. Of course, values of the inner
disk radius inferred from any continuum fits are suspect, and must be
viewed with caution.  A number of effects \citep{merloni00} can serve
to distort the observed inner disk parameters (see \citet{saito07} 
for a determination of more realistic values for the radii, since
these are not affected by a strong power-law component). According to
\citet{merloni00}, the dominant effect seems to be that the opacity is
dominated by electron scattering rather than  free-free absorption.
The net result is that the derived temperature given by the ${\rm
kT}_{in}$ parameter overestimates the effective inner temperature by a
factor of 1.7 or more \citep{shimura95}. 

The most interesting issue regarding our INTEGRAL observations of
GRO~J1655$-$40 is undoubtely the presence of  very significant and
unbroken high energy emission up to $500$\,keV in the hard state
of GRO~J1655-40, as noticed from first
epoch. \citet{grove98b} made a comparative measurement of a number  of
systems with CGRO/OSSE and found that these systems showed a  cut-off
at high energies at around ${\sim}100$\,keV in the hard state
of many systems \footnote{CGRO/OSSE
integration times were very long (of order of weeks), so different
states would be mixed. This is not an issue for INTEGRAL, since
exposures for each obtained spectrum are around two days and it is not
hoped to have noticeable high energy evolution with this timing, as
infrerred from the  light curves.}. Since then several studies have
tried to measure the cut-off in the spectra that could manifest the
validity of Comptonization processes by thermal electrons playing 
the major role in the high energy emission of black holes in the 
hard state. In our work we did not find such a break and the
presence of a non-thermal population of relativistic electrons was infrerred 
from the fitting of our spectrum in the hard state. 

Our finding is that non-thermal processes are the most important in order to
explain the high energy emission of the hard state of
GRO~J1655-40. Additionally, there is not any break in the data indicating 
that the high energy emission is mainly produced by thermal 
Comptonization as previously claimed. Moreover, non-thermal Comptonization is the
main source of the high energy emission in the second epoch
(thermal dominant state). This condition 
for the thermal dominant state has been broadly
reported in the literature since \citet{grove98b}.

In Table \ref{comp_previous} we summarize a list of references 
showing breaks in the high energy emission of several black hole
candidates in the hard and the intermediate state. One 
contemporaneous study by \citet{shaposhnikov06}
in the hard state of GRO~J1655-40 used
INTEGRAL observations covering a period of time slightly prior to that
of our first epoch. They pointed out the presence of a cut-off in
their data at around 200\,keV in their ISGRI and SPI spectra,
as a manifestation of the role of thermal Comptonization being the
main source of the high energy emission. However, their sensitivity at
energies ${\gtrsim}200$\,keV is not high enough in order to
disentangle the energy emission of a non-thermal population of
electrons. As also noticed by \citet{coppi99}, if a spectrum extends up
to $500$\,keV high energy emission coming from a non-thermal
population  of electrons is clearly present. The finding of
the contribution of the non-thermal population of electrons in the
high energy emission was reported before for the  Cyg X-1 system
(\citet{malzac06}, \citet{cadolle06} both in the intermediate  state)
and may be for the GRS~1915+105 (\citet{zdziarski01}, 
\citet{rodriguez04} both in the thermal dominant state). With our study, we
extend  the list of sources showing high energy emission coming from
non-thermal electrons with GRO~J1655-40, this time in the hard
state. \citet{joinet06} also reported the detection of a non-thermal
population of electrons in the hard state of GX 339-4. However, 
they indicated also the presence of a cut-off in the spectra. 

In Table \ref{comp_previous} it can be seen that while some cut-offs
appear to be close to the upper boundaries of the high energy instruments
used, others appear to be physically meaningful. The last correspond to
the systems GX 339-4 and Cyg X-1, both systems with known low inclination 
angle of ${\rm i}{\approx}45^{\circ}$ and ${\rm i}{\approx}35^{\circ}$,
respectively. GRO~J1655-40 is a very high inclination system with an inner
disk that could have an inclination of ${\rm i}=85^{\circ}$ (see Section
\ref{introd}). Comptonization processes of soft photons through thermal 
and/or non-thermal electrons seems to be highly isotropical from an
observer external to the system, if the electrons do not acquire outflow
velocities $v/c{\gtrsim}0.2$, \citep{beloborodov98}, so in principle it would
not depend on the inclination angle of the system. But other processes,
like reflection, could depend on the view angle, being almost unobserved for 
high inclination systems like GRO~J1655 -40 (as pointed out above with our fitting of epoch 3).  
So, GRO~J1655-40 in the hard state would be an excellent source for the studying of 
Comptonization processes, without the apparent disturbance of any other 
high energy mechanisms (including soft emission from a disk). If these hypothesis 
are correct, we would have discovered that Comptonization of black 
hole transients in the hard
state occurs by mainly Comptonization through non-thermal relativistic electrons.
This conclusion is also supported by the studies of \citet{malzac06} of 
Cyg X-1 in the intermediate state because while showing a high energy cut-off at 
the energy around 100\,keV, they also point out to the presence of high 
energy emission from a non-thermal population of electrons in an state close
to the hard (in the case of \citet{joinet06} in the hard state of
GX 339-4).

\acknowledgments

We thank Danny Steeghs for helpful discussions. We also thank the
RXTE instrument teams at MIT and NASA/GSFC for providing the ASM light
curve.  MDCG is a MEC funded PhD student supported under grants
PNE2003-04352+ESP2005-07714-C03-03.  This work is based on
observations made with INTEGRAL, an ESA science mission with
instruments and science data centre funded by ESA member states and
with the participation of Russia and the USA. We thank to K. Arnaud
for providing part of the eqpair code used in this work. We finally thank the
anonymous referee for the careful reading of this manuscript.




\clearpage

\begin{deluxetable}{cccccc}
\rotate
 \tablecaption{INTEGRAL observing ToO during which GRO J1655-40 was observed in the 2005 outburst, giving the exposure times of the summed spectra analyzed for each ToO and instrument.\label{tobserv}}
 \tablewidth{0pt}
 \tablehead{\colhead{Epoch number} & \colhead{INTEGRAL revolution \& MJD} & \colhead{Start \& End Date} & \colhead{JEM-X} & \colhead{ISGRI} & \colhead{SPI}
 }
 \startdata
           &       (days)        & (yyyy/mm/dd)   &  [s]  &   [s]      & [s] \\
           &                     &                &       &            &      \\
\hline
         1 & 290  & 2005/02/27-28 & 44504.83  & 69212.87 & 90849.58 \\
           &  53428.20-53429.50    &                       &           &         & \\
         2 & 295-296 & 2005/03/16-18 & 43641.93  & 71827.16 & 92584.05 \\
           &  53445.10-53447.80    &                       &           &          &          \\
         3 & 299 & 2005/03/26-28 & 44673.42  & 69848.25 & $-$  \\
           &  53455.80-53457.00    &                       &           &          &          \\
         4 & 304 & 2005/04/10-11 & 46037.85  & 67883.58 & $-$  \\
           &  53470.00-53471.33    &                       &    &   &          \\
 \enddata
 \end{deluxetable}

\begin{deluxetable}{ccccc}
\tabletypesize{\scriptsize}
\rotate
\tablecaption{Parameters obtained for the best fits of the joint JEM-X, ISGRI and SPI spectra (see text for details), using the model $constant{\times}eqpair$ 
in the first epoch, $constant{\times}phabs(diskbb+eqpair)$ in the second epoch, 
$constant{\times}phabs(diskbb + gaussian + pexriv)edge{\times}edge$ in the third epoch and $constant{\times}phabs(diskbb + gaussian + powerlaw)edge{\times}edge$
in the fourth epoch.\label{param_spec}}
\tablewidth{0pt}
\tablehead{
\colhead{} & \colhead{{\rm Epoch 1}} & \colhead{{\rm Epoch 2}} & \colhead{{\rm Epoch 3}} & \colhead{{\rm Epoch 4}}
}
\startdata
Parameter    &                         &           &           &                 \\
&                                &           &           &                 \\
&                                     &  Powerlaw &           &                 \\
&                                  &           &           &                 \\
${\Gamma}$   &     $-$                 & $-$          &    $-$    &    $4.7{\pm}0.6$ \\
${\rm N}_{\rm pow}$ [${\rm ph}/{\rm keV}/{\rm s}/{\rm cm}^{2}$] at 1 keV &  $-$              &  $-$     &  $-$      &  $580{\pm}60$   \\
&                                  &           &           &                 \\
&                                     &  eqpair &           &                 \\
&                                  &           &           &                 \\
 ${\Gamma}_{\rm p}$    &       $0.6{\pm}0.3$     &   ${\approx}0$       &  $-$      &     $-$         \\
 ${\gamma}_{\rm min}$       &    $1.3$ (f)            &  $1.3$ (f)           &  $-$      &     $-$         \\
 ${\gamma}_{\rm max}$       &   $1000$ (f)            &  $1000$ (f)          &  $-$      &     $-$         \\
 ${\rm l}_{s}$              &     $10$ (f)            &  $10$ (f)            &  $-$      &   $-$           \\
 ${\rm l}_{h}/{\rm l}_{s}$   &    $0.8{\pm}0.3$        &  $1.2{\pm}0.6$       &  $-$      &    $-$          \\
 ${\rm l}_{nth}/{\rm l}_{h}$  &   $0.8{\pm}0.1$         &  $0.2{\pm}0.3$       &   $-$     &    $-$          \\
${\tau}$                    &   $1{\pm}0.5$     &    $4{\pm}1$        &   $-$              &    $-$                \\
${\rm Refl}$ $[{\Omega}/2{\pi}]$ & 0(f)          & $0$ (f)       &     $-$       &     $-$       \\
${\rm kT}_{\rm in}$\,(keV)    &    $0.5$(f)        & $1.25{\pm}0.02$   &  $-$      &    $-$          \\
${\rm R}_{\rm in}$\,(keV) [${\rm GM/c}^{2}$]    &    $6$ (f)                       &    $6$ (f)        &  $-$      &    $-$          \\
${\rm R}_{\rm out}$\,(keV) [${\rm GM}/c^{2}$]    &   $100$ (f)                      &   $100$ (f)       &  $-$      &    $-$          \\
&                                  &           &           &                 \\
&                                     &  diskbb   &           &                 \\
&                                  &           &           &                 \\
${\rm kT}_{\rm in}$ [keV] &       $-$        &  $1.25{\pm}0.02$         &  $1.28{\pm}0.02$         & $1.27{\pm}0.17$               \\
${\rm N}_{\rm bb}$ [$({\rm R}_{\rm in}[{\rm km}]/{\rm D}[10{\rm kpc}])^{2}{\times}{\rm cos}{\theta}$]        &       $-$                   &  $1321^{+72}_{-67}$   & $710{\pm}40$   & $500{\pm}60$  \\
&                                  &           &           &                 \\
&                                     &  gaussian &           &                 \\
${\rm E}_{\rm gauss}$ [keV] &     $-$            &    $-$    & $6.7{\pm}0.3$   &   $6.7{\pm}0.9$  \\
${\sigma}$ [keV] &      $-$            &    $-$    & $0.63{\pm}0.15$ &  $0.8{\pm}0.5$     \\
${\rm N}_{\rm gauss}$ [${\rm ph}/{\rm cm}^{2}/{\rm s}$]    &    $-$    &    $-$    &  $0.017{\pm}0.008$  &   $0.05{\pm}0.03$     \\
&                                     &  edge (Fe XXV)    &           &          \\
${\rm E}_{\rm edge}$ [keV] &      $-$            &   $-$     & $8.64{\pm}0.20$  &    $8.6{\pm}0.9$ \\
${\tau}$ &              $-$            &   $-$     & $0.20{\pm}0.05$ &  $0.20{\pm}0.10$  \\
&                                     &  edge (Fe XXVI)    &           &          \\
${\rm E}_{\rm edge}$ [keV] &      $-$            &   $-$     & 9.278 (f) &  9.278 (f)      \\
${\tau}$ &              $-$            &   $-$     & ${\leq}0.02$ (f)  &   ${\leq}0.02$ (f)    \\
&                                     &  pexriv   &           &                 \\
Photon index &          $-$            &   $-$     & $2.50{\pm}0.23$ &  $-$ \\
${\rm E}_{\rm f}$ [keV] &         $-$            &   $-$     & 1000.0 (f)&   $-$           \\
${\rm R}$ $[{\Omega}/2{\pi}]$ &      $-$     &   $-$     & ${\leq}0.30$  &   $-$         \\
${\rm Fe}$ abundance &        $-$            &   $-$     & 2.8 (f)   &   $-$          \\
${\rm cos(i)}$  &             $-$            &   $-$     & 0.45 (f)  &    $-$          \\
${\rm T}_{\rm disk}$ [K] &        $-$            &   $-$     & 1.2E+07 (f) &    $-$         \\
${\rm xi}$ [$4{\pi}{\rm F}_{\rm irr}/{\rm n}$] &   $-$       &   $-$     & 5000 (f)  &    $-$ \\
${\rm N}_{\rm pexriv}$ [${\rm ph}/{\rm keV}/{\rm cm}^{2}/{\rm s}$]  &   $-$   &   $-$     &   $3.8{\pm}1.5$  &  $-$              \\
&                                     &           &           &                 \\
&                                     &           &           &                 \\
&                                     &           &           &                 \\
&                                     &           &           &                 \\
&                                     &           &           &                 \\
&                                     &  Intrumental normalization factors &           &                 \\
&                                     &           &           &                 \\
${\rm C}_{\rm JEM-X}$  &   1.0 (f)               &  1.0 (f)  &  1.0 (f)  &     1.0 (f)   \\
${\rm C}_{\rm ISGRI}$  &  $1.1{\pm}0.1$                 &  $1.0{\pm}0.1$    & $0.46^{+0.11}_{-0.15}$ &  $0.39{\pm}0.14$  \\
${\rm C}_{\rm SPI}$  &   $1.4{\pm}0.1$             &  $1.1{\pm}0.1$   & $-$     &     $-$         \\
&                                     &           &           &                 \\
${\chi}_{\nu}^{2}$ &   1.21            &  1.44     &  2.62     &   2.73          \\
${\nu}$ &              57              &  83       &  19       &   22            \\
&                                    &           &           &                 \\
\enddata
\tablecomments{Parameters fixed in the fits are denoted by 'f'. We fixed the value of the column density to $N_{H}=8.0{\times}10^{21}$ atoms$/cm^{2}$ as obtained by \citet{diaz06} using XMM-Newton data of GRO~J1655$-$40 during the same outburst.}

\end{deluxetable}

\clearpage
\thispagestyle{empty}
\setlength{\voffset}{20mm}
\begin{deluxetable}{cccccccc}
\rotate
 \tablecaption{Values for the components of Comptonization models in previous studies of several sources in several states close to the hard (LH means hard 
state and IS means intermediate state). Also, values for the high-energy cut-offs and/or
break energies (if present) are reported.\label{comp_previous}}
 \tablewidth{0pt}
 \tablehead{\colhead{${\Gamma}$} & \colhead{${\rm State}$} & \colhead{$E_{cut-off}$(\,keV)} & \colhead{$E_{max}$(\,keV)} &  $T_{e}$ (\,keV) & ${\tau}$ & \colhead{Source} & \colhead{Reference}
 }
 \startdata
$1.67{\pm}0.06$ &   LH    &  $195{\pm}50$  &  ${\approx}200$ &  $35^{+200}_{-9}$ & $1.45^{+0.5}_{-1.4}$  &  IGR~J17497$-$2821 & Walter et al. 2007, A\&A, 461, L17 \\
$1.4-1.6$       &   LH    & $50-200$       &  ${\approx}200$ &  $-$              &   $-$    &   GX~339$-$4       &  Miyakawa et al. 2007, astro-ph/0702087 \\
$1.70{\pm}0.01$ &   LH    &  $115{\pm}5.6$ &                 &  $-$              &  $-$     &  XTE~J1550$-$564   &  Yuan et al. 2006, astro-ph/0608552 \\
$1.53{\pm}01$   &  LH     &  $460{\pm}300$ &                 &  $-$              &  $-$     &  XTE~J1550$-$564   &  Yuan et al. 2006, astro-ph/0608552 \\
$1.92{\pm}0.05$ &  IS     &   $72{\pm}8$   & ${\approx}200$  &   $-$             &   $-$    &  GX~339$-$4        &  Belloni et al. 2006, MNRAS, 367, 1113 \\
$1.9{\pm}0.1$   &  LH     &${\approx}150$  & ${\approx}600$  &  $67^{+8}_{-6}$   & $1.98{\pm}0.22$ &  Cyg X-1    &  Cadolle Bell et al. 2006, A\&A, 446, 591 \\
$2.1$           &  IS     &${\approx}100$  & ${\approx}1000$ &  $20-65$          & $0.55-1.36$ &  Cyg X-1        &  Malzac et al., 2006, A\&A, 448, 1125 \\
$1.65-2.0$      &  LS     &  $130-250$     & ${\approx}150$  &  $-$              &   $-$    &  Cyg~X$-$1         &  Wilms et al. 2006, A\&A, 447, 245 \\
$1.35{\pm}0.03$ &  LH     &  $100-200$     & ${\approx}600$  &                   &          & GRO~J1655$-$40     &  Shaposhnikov et al.,2006, astro-ph/0609757 \\
$1.72{\pm}0.03$ &  LH     &    $-$         & ${\approx}500$  &  $-$               &         & GRO~J1655$-$40     & This work \\
 \enddata
 \end{deluxetable}
\clearpage
\setlength{\voffset}{0mm}




\setcounter{figure}{0}

\begin{figure}
\centering
\includegraphics[width=1.0\linewidth]{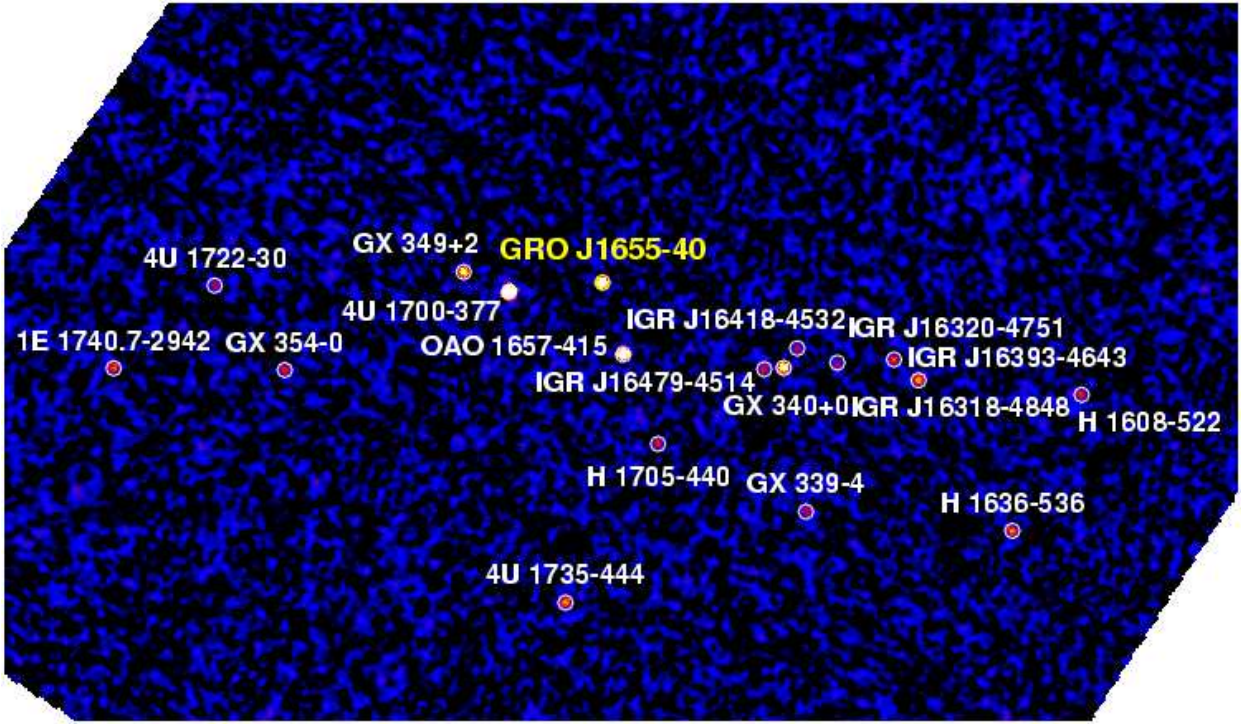}
\caption{Mosaic significance image (obtained in revolution 295) of the GRO~J1655-40 region as seen with ISGRI in the $20-40$\,keV energy range. Besides the target source, several other high energy sources are visible.}
\label{mosaic}
\end{figure}

\clearpage
\begin{figure}
\centering
\includegraphics[width=0.6\linewidth]{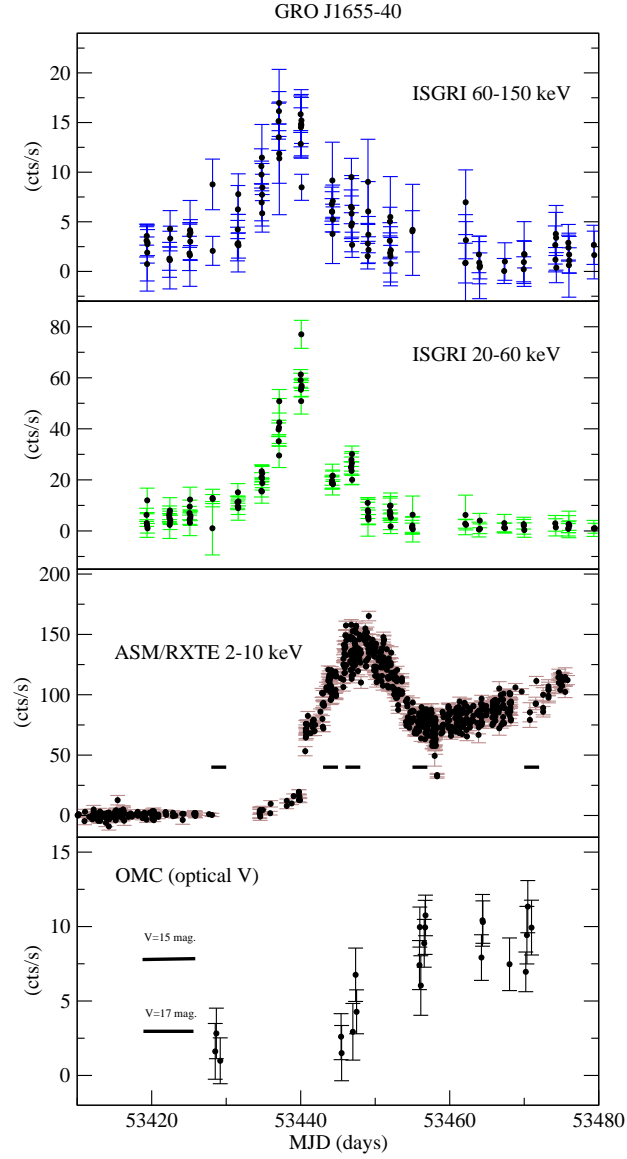}
\caption{Light curves obtained with ISGRI from the INTEGRAL Galactic Monitoring Program in two energy
bands (60-150\,keV and 20-60\,keV), together with OMC (optical). The horizontal lines in the OMC panel show the equivalence in magnitudes of the fluxes. In the third panel, ASM/RXTE (2-12\,keV) light curve is shown in the same period of time. The horizontal
lines indicate the time intervals (one revolution each) over which INTEGRAL spectra were obtained.}
\label{lcurves}
\end{figure}

\clearpage
\begin{figure}
\centering
\includegraphics[width=0.6\linewidth]{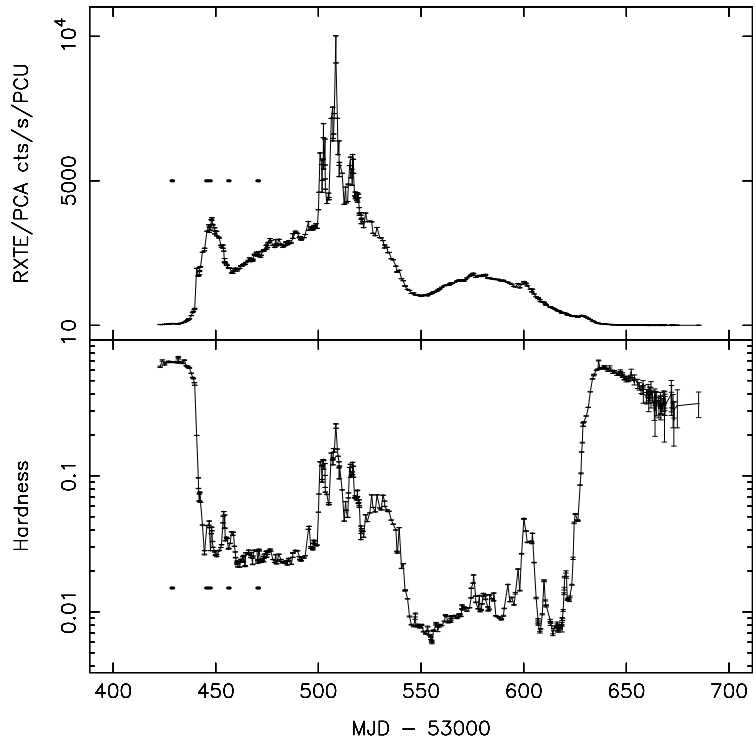}
\caption{RXTE/PCA light curves in the 2-60\,keV band and hardness ratio calculated as the ratio of
the 9.4-18.2\, keV and 2.8-5.7\,keV count rates. The curves were made from 520
RXTE observations, with one (averaged) data point from each observation. Data taken from \citet{homan07}.
}
\label{lcurves2}
\end{figure}

\begin{figure*}
\centering
\includegraphics[angle=270,width=1.0\linewidth]{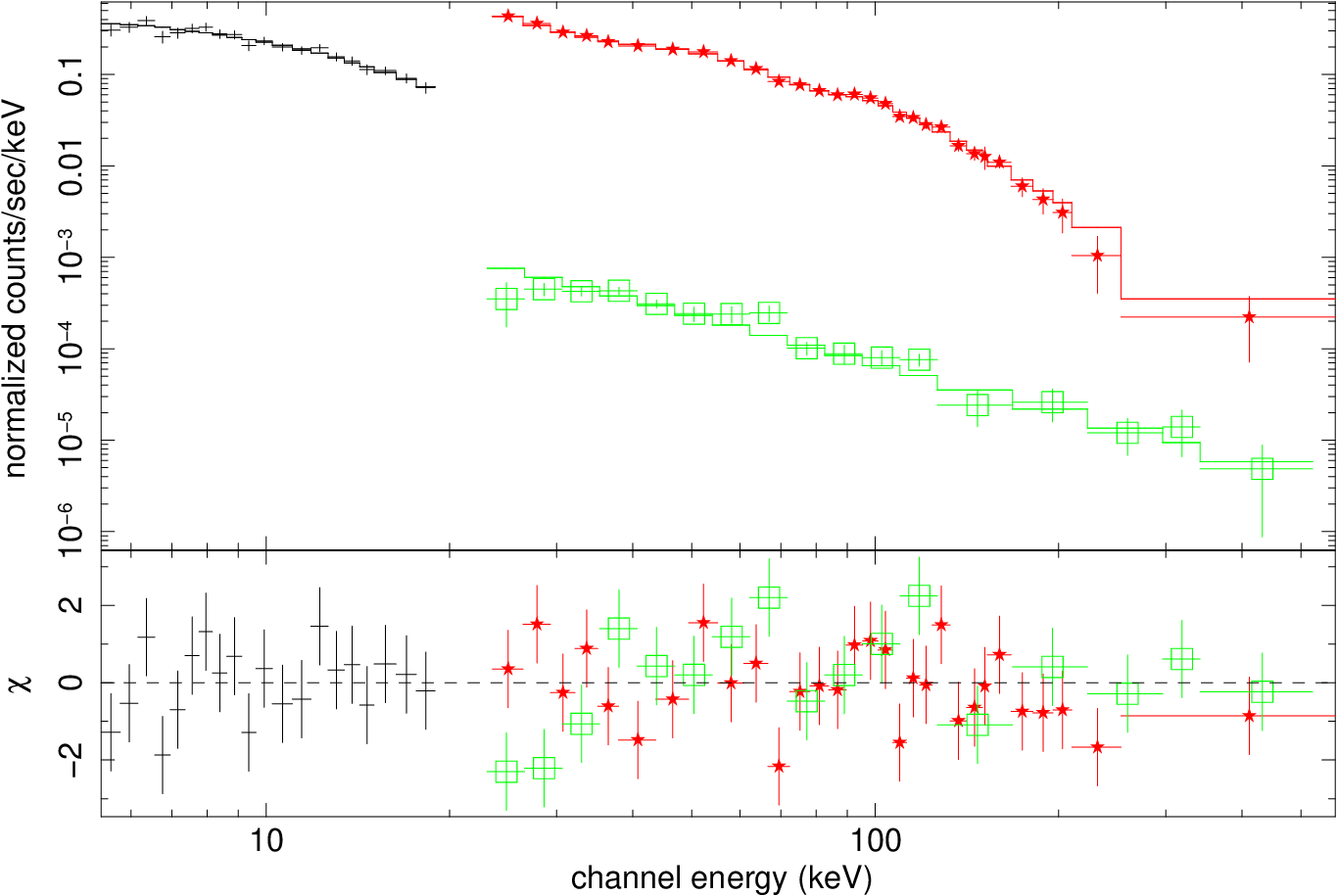}
\caption{Fitted INTEGRAL spectra, corresponding to epoch 1 with a simple phenomonological power-law model. Refer to the text in Section \ref{powerlaw} and \ref{comptteqpair} in order to get detailed information about this fit.}
\label{fspecpo1}
\end{figure*}

\begin{figure*}
\centering
\includegraphics[angle=270,width=1.0\linewidth]{f5.eps}
\caption{Fitted INTEGRAL spectra, corresponding to epoch 2 with a simple phenomonological power-law model. Refer to the text in Section \ref{powerlaw} and \ref{comptteqpair} in order to get detailed information about this fit.}
\label{fspecpo2}
\end{figure*}

\clearpage
\begin{figure}
\centering
\includegraphics[angle=270,width=1.0\linewidth]{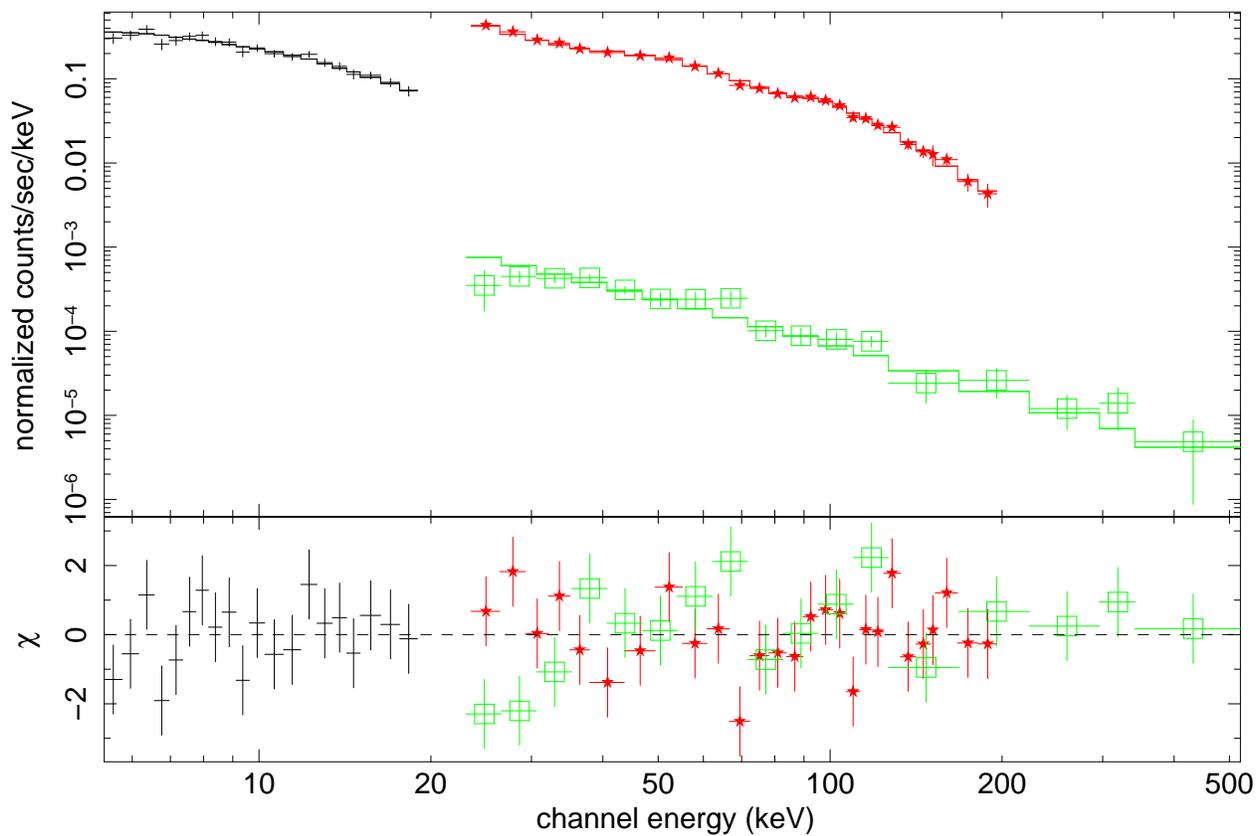}
\caption{Fitted INTEGRAL spectra, corresponding to epoch 1 (hard state). These spectra were fitted with the EQPAIR Comptonization model of \citet{coppi99} ($eqpair$ 
model in XSPEC). Details about the fitting and the parameters obtained in Section \ref{compttmod} and in Table \ref{param_spec}. JEM-X (single line -black-), ISGRI 
(star line -red-) and SPI spectra (square line -green-), are shown, respectively.}
\label{fspec1}
\end{figure}

\clearpage
\begin{figure}
\centering
\includegraphics[angle=270,width=1.0\linewidth]{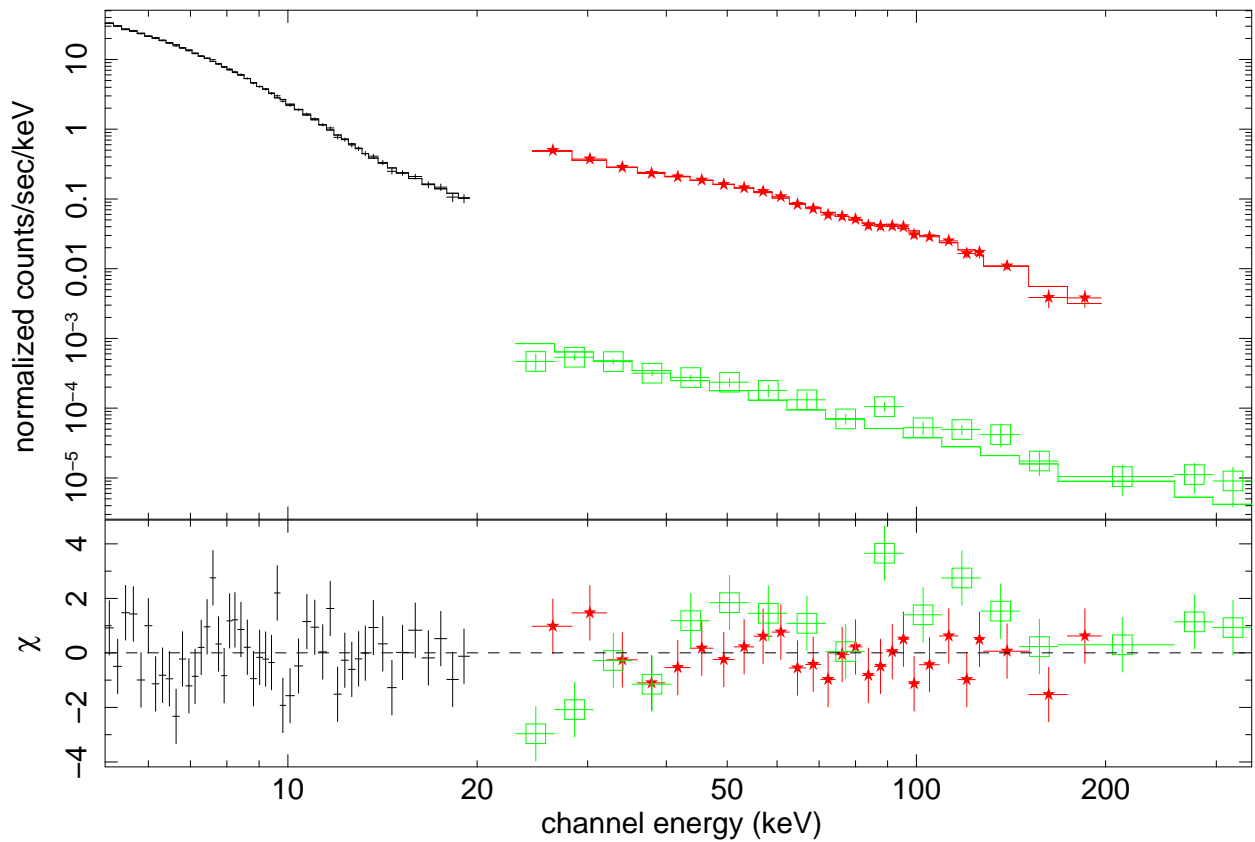}
\caption{Fitted INTEGRAL spectra, corresponding to epoch 2 (thermal dominant state). These spectra were fitted with the EQPAIR Comptonization model of \citet{coppi99} ($eqpair$ 
model in XSPEC) considering also the soft emission from an accretion disk ($diskbb$ model in XSPEC). Details about the fitting and the parameters obtained in 
Section \ref{compttmod} and in Table \ref{param_spec}. JEM-X (single line -black-), ISGRI (star line -red-) and SPI spectra (square line -green-), are shown, respectively.}
\label{fspec2}
\end{figure}

\clearpage
\begin{figure}
\centering
\includegraphics[angle=270,width=1.0\linewidth]{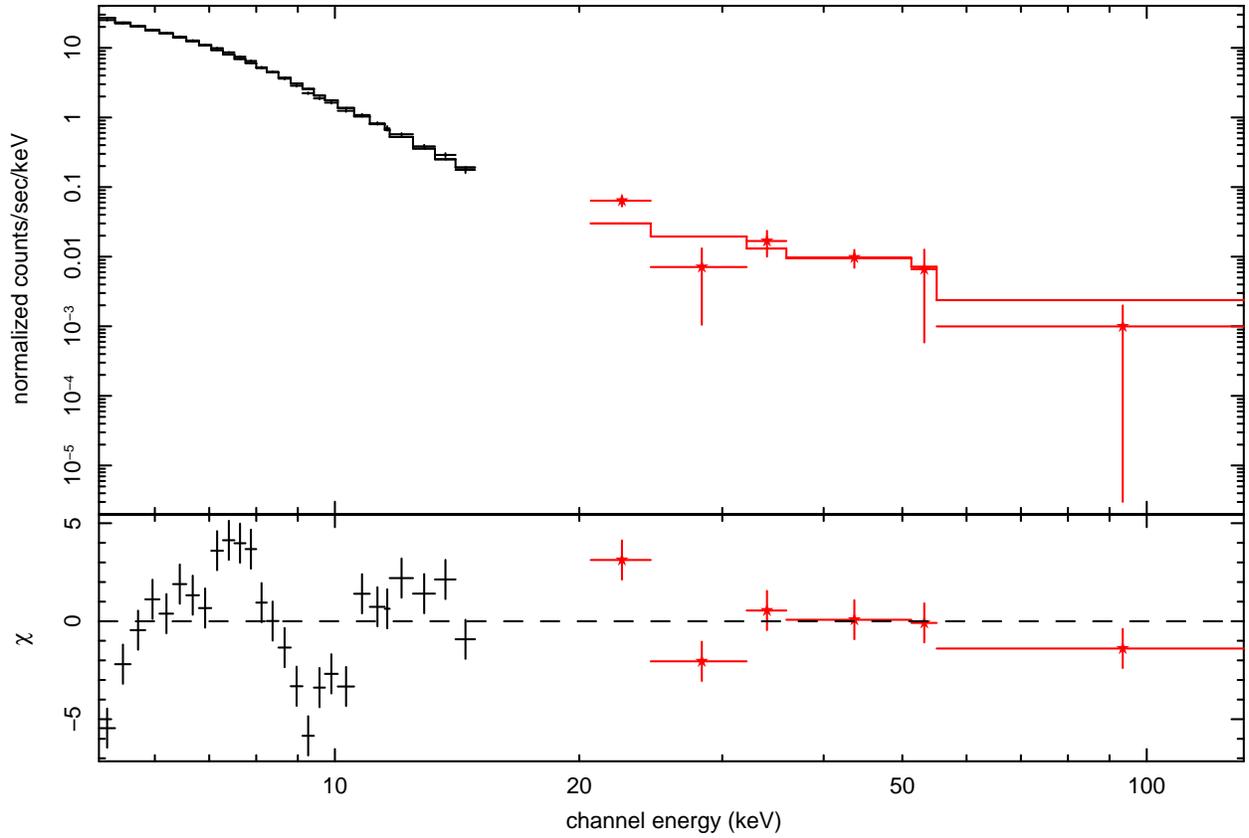}
\caption{Fitted INTEGRAL spectra, corresponding to epoch 4 (thermal dominant state). The model used was one consisted of a pure power-law plus emission from an absorbed
multicolor accretion disk of \citet{mitsuda84}. In this period we noticed a change in the properties of the accretion outflow with respect to epochs 1 and 2 (see 
Section \ref{powerlaw} and \ref{discuss} for details).}
\label{fspec3}
\end{figure}

\clearpage
\begin{figure}
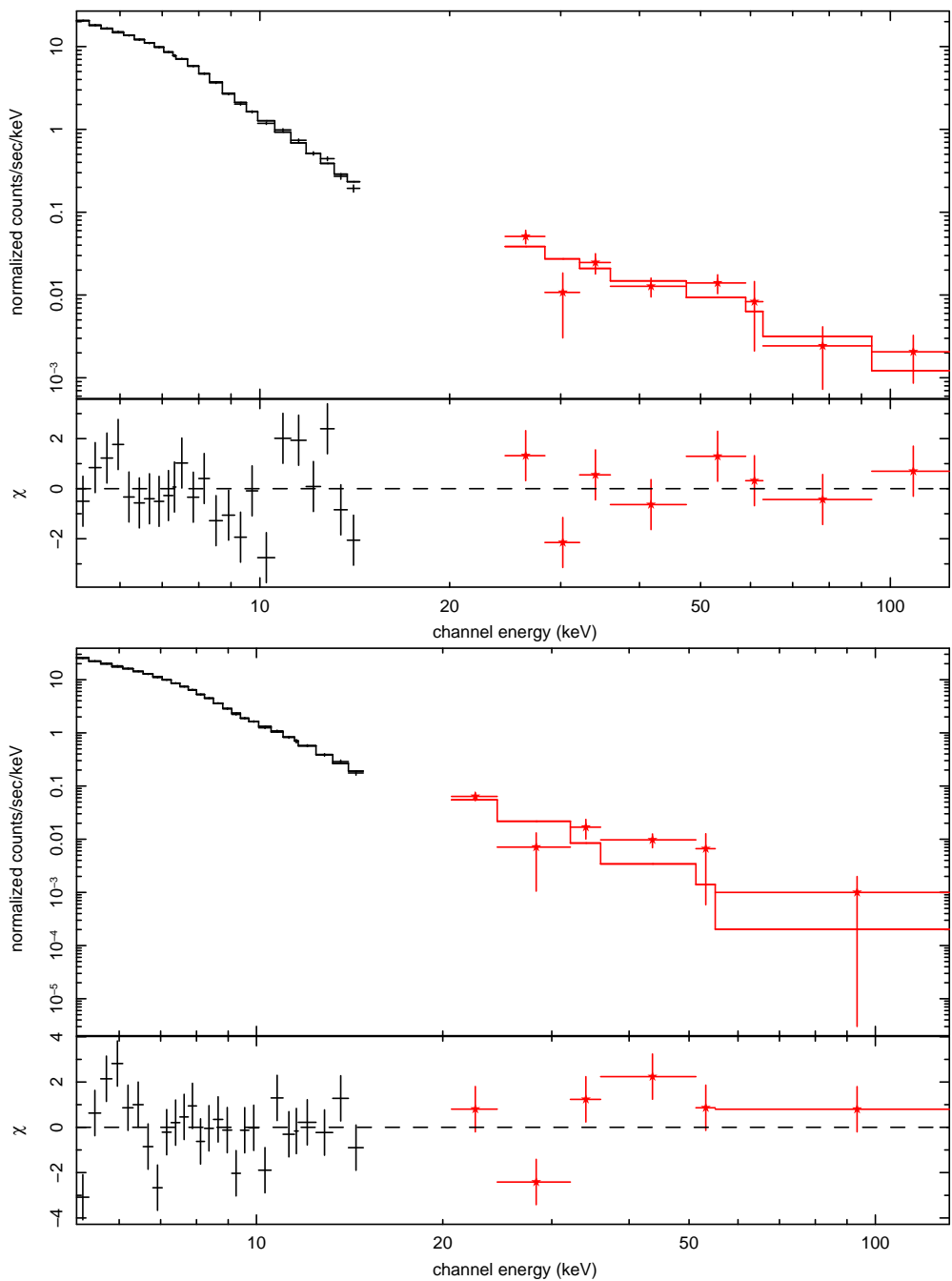

\centering
\includegraphics[angle=270,width=0.8\linewidth]{f9a.eps}
\includegraphics[angle=270,width=0.8\linewidth]{f9b.eps}
\caption{Fitted INTEGRAL spectra, corresponding to epochs from 3 and 4 (upper to lower panel). These spectra were fitted with an absorbed multicolor disk 
\citep{mitsuda84} plus a simple power-law iron emission line and edges for both two epochs (and reflection in the case of epoch 3). Details about the fitting
and the values of the parameters obtained in Section \ref{edges} and Table \ref{param_spec}. JEM-X (single line -black-) and ISGRI (star line -red-) are shown, 
respectively.}
\label{fspec4}
\end{figure}

\clearpage
\begin{figure}
\centering
\includegraphics[angle=270,width=0.4\linewidth]{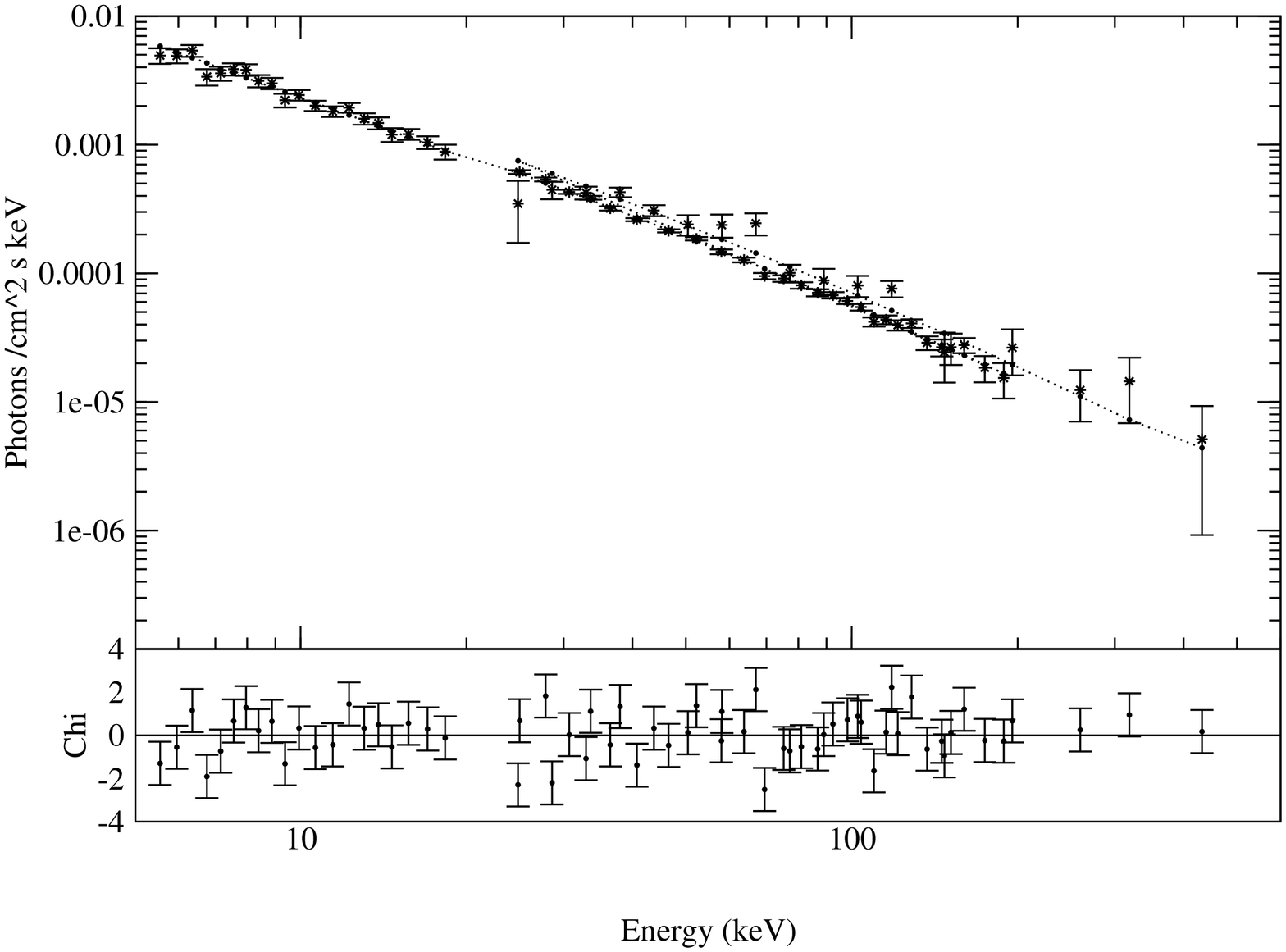}
\includegraphics[angle=270,width=0.4\linewidth]{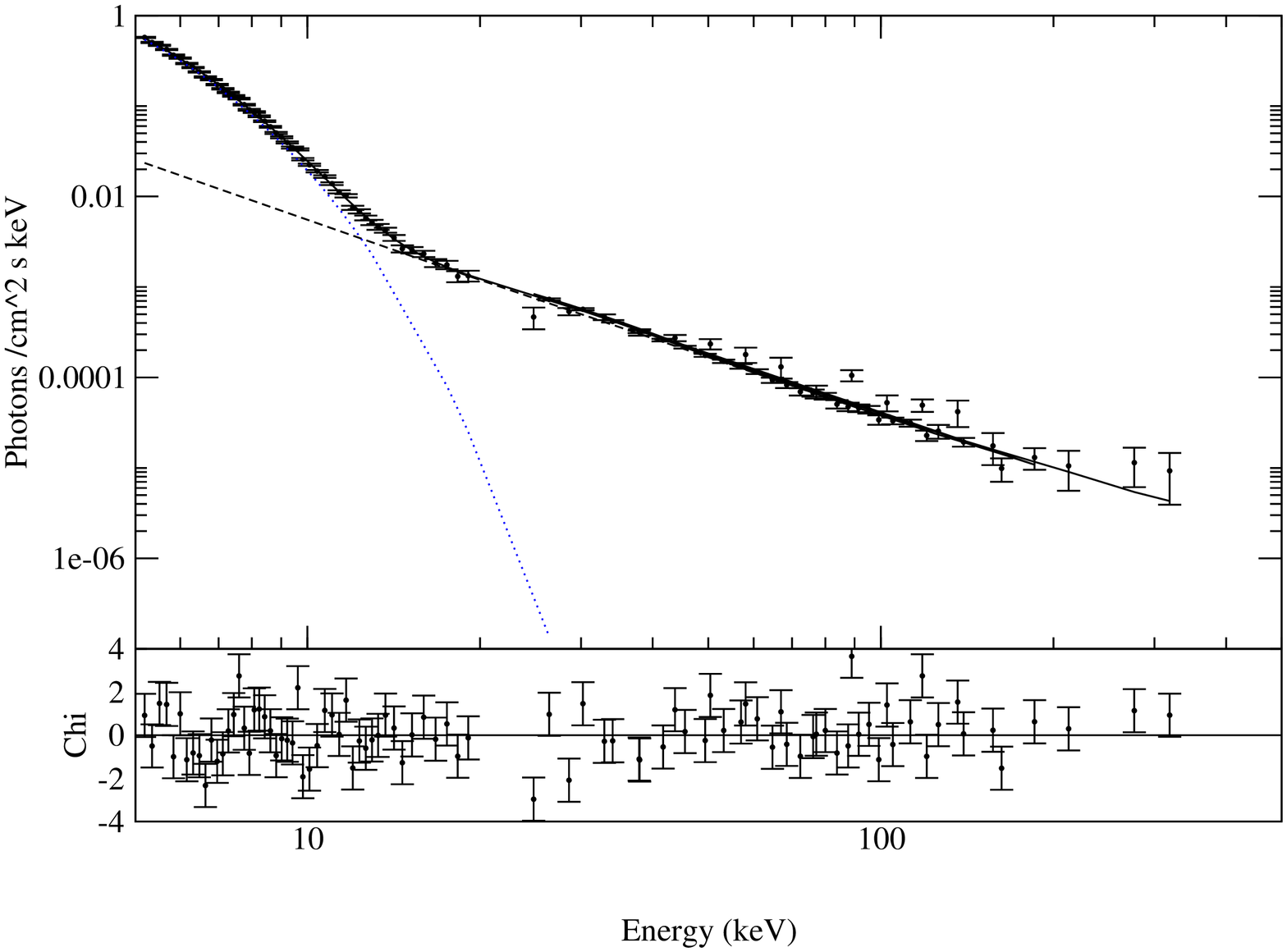}
\includegraphics[angle=270,width=0.4\linewidth]{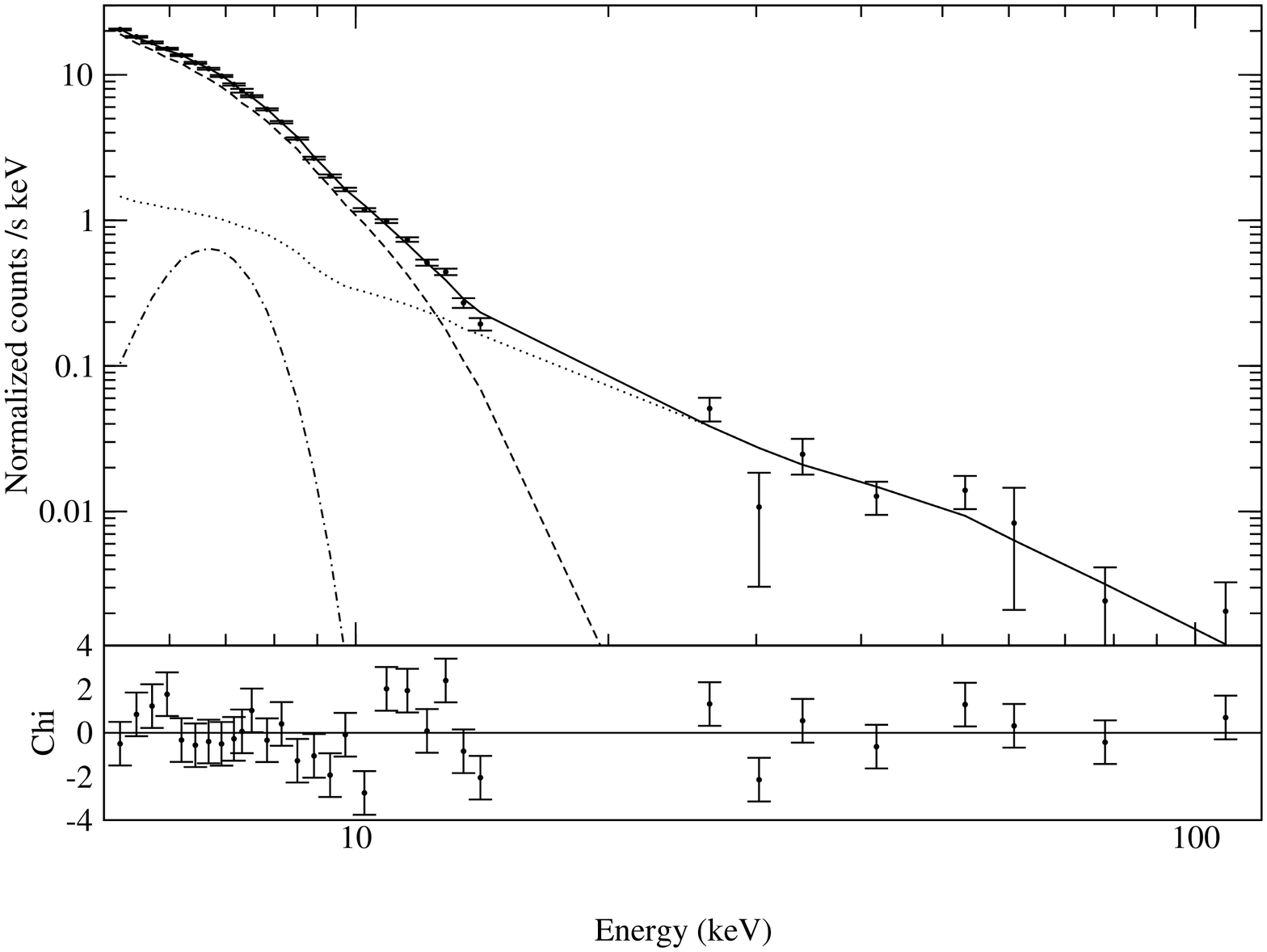}
\includegraphics[angle=270,width=0.4\linewidth]{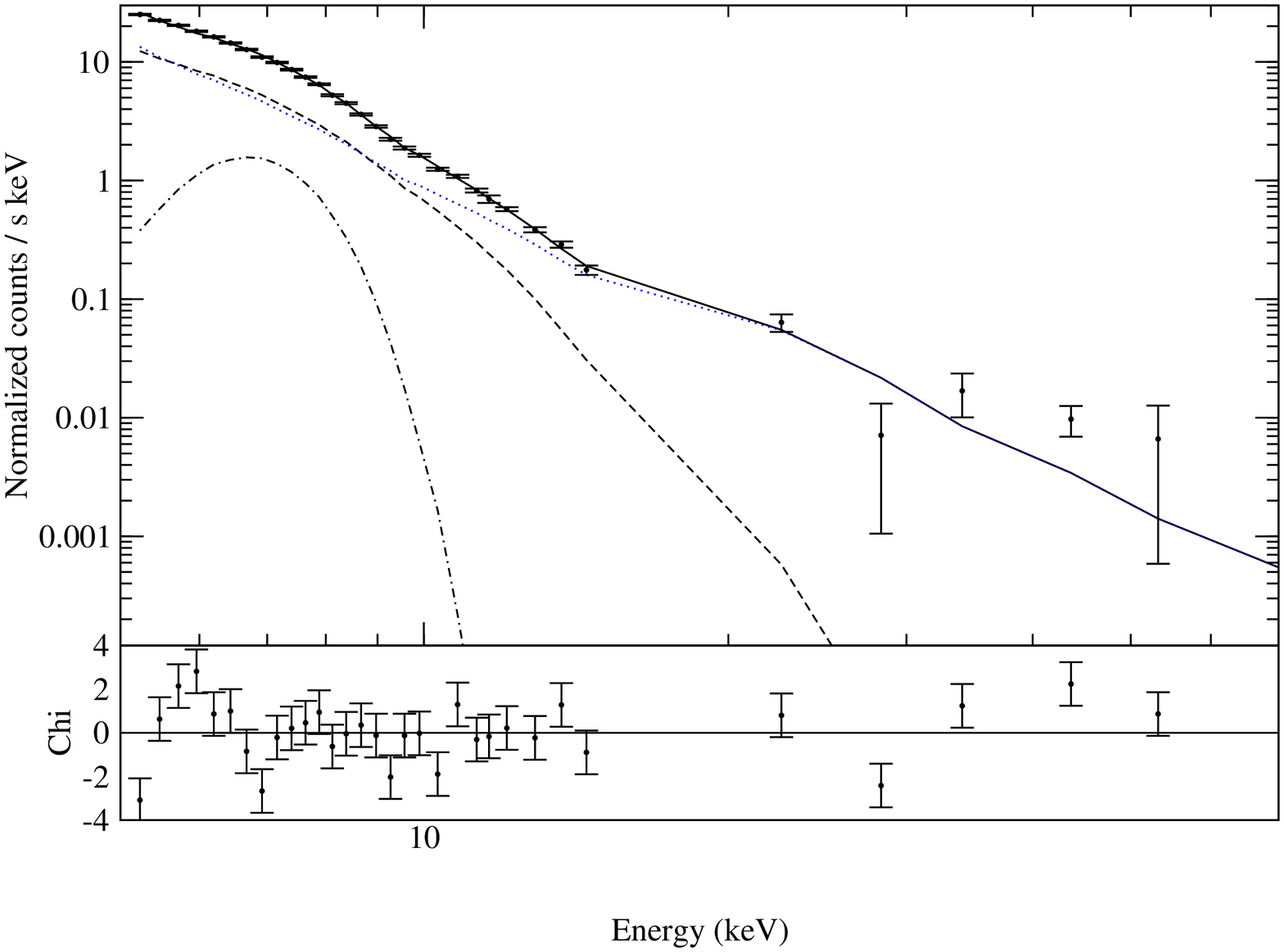}
\caption{Unfolded spectra from epochs 1 to 4 (upper-left to lower-right). The continuum line shows the total model (see the text and Table \ref{param_spec} for 
details), the dashed-dotted line shows the iron K${\alpha}$ emission line, the long dashed line the accretion disk component and the short dashed line the 
power-law ($pexriv$ in the third epoch).}
\label{fspec5}
\end{figure}

\end{document}